\documentclass[prb,floatfix,amsmath,amssymb,superscriptaddress]{revtex4}
\usepackage{graphicx}
\usepackage{latexsym}
\usepackage{graphicx}
\usepackage{times}
\usepackage{color}
\usepackage{amsmath}
\usepackage{dcolumn}
\usepackage{latexsym,amsmath,amssymb,bm,euscript}
\newcommand{\ket}[1]{\left|{#1}\right\rangle}
\begin{document} 

\title{Topology driven quantum phase transitions in time-reversal invariant anyonic quantum liquids} 

\author{Charlotte Gils}
\affiliation{Theoretische Physik, ETH Zurich, 8093 Zurich, Switzerland}
\author{Simon Trebst}
\affiliation{Microsoft Research, Station Q, University of California, Santa Barbara, 
CA 93106}
\author{Alexei Kitaev} 
\affiliation{Institute for Quantum Information, California Institute of Technology, 
Pasadena, CA 91125}
\author{Andreas W. W. Ludwig}
\affiliation{Physics Department, University of California, Santa Barbara, CA 93106}
\author{Matthias Troyer}
\affiliation{Theoretische Physik, ETH Zurich, 8093 Zurich, Switzerland}
\author{Zhenghan Wang}
\affiliation{Microsoft Research, Station Q, University of California, Santa Barbara, 
CA 93106}

\date{\today} 

\begin{abstract} 
\bf
Indistinguishable particles in two dimensions can be characterized by anyonic
quantum statistics more general than those of bosons or fermions. 
Such anyons emerge as quasiparticles in fractional quantum Hall states 
and certain frustrated quantum magnets.
Quantum liquids of anyons exhibit degenerate ground states where the degeneracy 
depends on the topology of the underlying surface.
Here we present a novel type of continuous quantum phase transition
in such anyonic quantum liquids that is driven by quantum fluctuations of topology. 
The critical state connecting two anyonic liquids on surfaces with different topologies is 
reminiscent of the notion of a `quantum foam' with fluctuations on all length scales.
This exotic quantum phase transition arises in a microscopic model of interacting anyons
for which we present an exact solution in a linear geometry. 
We introduce an intuitive physical picture of this model that unifies string nets 
and loop gases,  
and provide a simple description of topological  quantum phases and their phase transitions.
\end{abstract} 

\maketitle 


\noindent
Phases of matter can exhibit a vast variety of ordered states that typically arise from
spontaneous symmetry breaking and can be described by a local order parameter.
A more elusive form of order known as `topological order' \cite{Wen89} reveals itself 
through the appearance of robust ground-state degeneracies, but cannot be described in terms 
of a local order parameter. 
Examples of such topological quantum liquids are the fractional quantum Hall states 
\cite{Laughlin83} where the ground-state degeneracy depends on the number of 
\lq antidots\rq \ 
which can be viewed as punctures (holes) in the two-dimensional
surface populated by the quantum Hall liquid  \cite{Wen_Niu90}.
It has long been proposed that topological quantum liquids also occur in certain 
frustrated quantum magnets
\cite{Moessner01,Balents01,Ioffe02,LevinWen05,Kitaev03,Kitaev06}, 
but it has only been in recent years that strong candidate materials
have emerged \cite{herbert,NaIrO}.
While quantum Hall liquids break time-reversal symmetry, the exotic ground states of
frustrated quantum magnets are expected to preserve time-reversal symmetry. 
As a consequence of this symmetry many unexplored phenomena may appear, including
the intriguing possibility of topology driven quantum phase transitions which is the 
central aspect of this manuscript.

\begin{figure}[b]
\includegraphics[width=0.95\columnwidth]{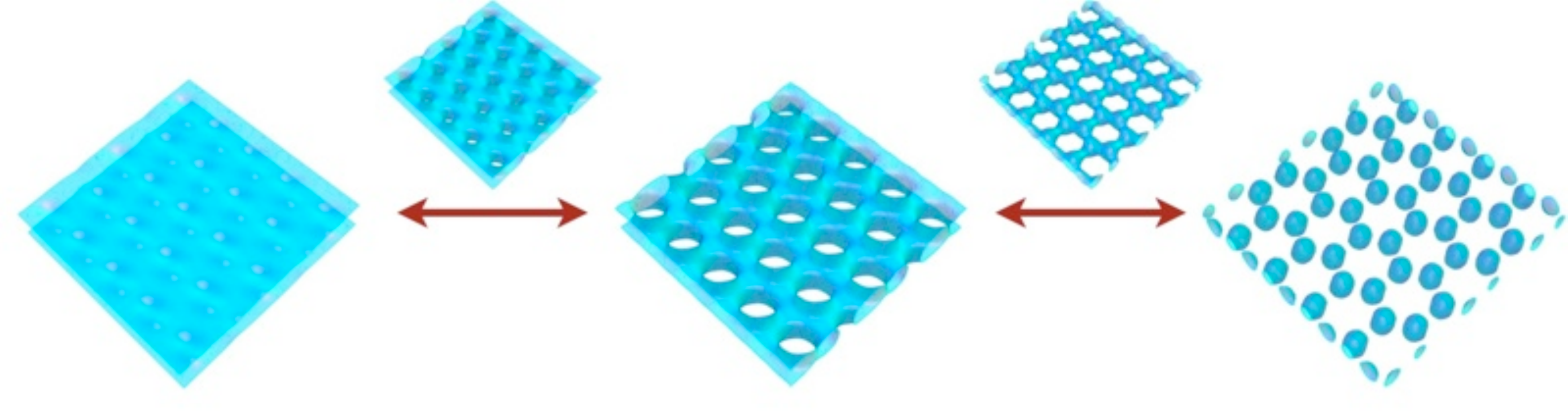}
\vskip -1mm
\caption{{\it Phase transition in two dimensions}. Two-dimensional surfaces with different topologies that are populated by anyonic 
                quantum liquids.
                A quantum phase transition driven by fluctuations of the surface topology 
                connects the anyonic liquid on two separated sheets (on the left) 
                and decoupled spheres (on the right).}
\label{Fig:Transition2D}
\end{figure}

In this manuscript we develop an intuitive physical picture for the emerging  low energy physics 
of topological quantum liquids and their phase transitions in terms of surfaces and their topology. 
We thereby provide a  
visualization of the underlying quantum physics, which 
is in one-to-one correspondence to a detailed analytical framework.
Here we consider systems that preserve time-reversal symmetry which in this picture will 
be described by quantum liquids on closed surfaces.
Such liquids exhibit ground-state degeneracies that depend (exponentially) on the genus of the surface. 
A section of an extended high-genus surface formed by a triangular arrangements of `holes' is 
shown in Fig.~\ref{Fig:Transition2D}. 
Through every such hole there can be a flux of the  
liquid populating the surface.
An exponential degeneracy then arises from the possible flux assignments through the holes.
While in the presence of a flux a hole cannot be contracted, we can eliminate the hole in the
absence of flux without changing the state of the topological liquid. 
If there is no flux through any of the holes, they can all be removed, and the state of the 
quantum liquid is identical to that on two separated sheets, as shown on the left side  in Fig.~\ref{Fig:Transition2D}. It is this state that exhibits topological order.
On the other hand, if there is no flux through the tubes
in the interior of the surface
(centered around the black lines in
Fig.~\ref{Fig:Model}), we
can pinch them off.  
The
resulting
 state
 of the quantum liquid is then identical to that of disconnected spheres, 
as shown on the right side in Fig.~\ref{Fig:Transition2D}.
This state has neither ground-state degeneracy  nor topological order.

Here we will introduce a microscopic  
model which 
energetically favors
the absence of flux through the holes or tubes,
thus dynamically implementing the two topology changing processes 
mentioned above. 
The competition of the two processes drives a quantum phase transition
between the  
two extreme states. 
Our model is defined on the `skeleton' that surrounds the holes in the 
interior of the surface as illustrated in Fig.~\ref{Fig:Model} where the skeleton 
forms a honeycomb lattice. 
The fluxes in the tubes are associated with discrete
degrees of freedom on the edges of the skeleton lattice, corresponding to anyonic 
particles of the quantum liquid \cite{Leinaas_Myrheim77}.
The set of degenerate ground states of the liquid is now 
in one-to-one correspondence with
all labelings of the edges
consistent with a given set of constraints, characteristic to the underlying quantum liquid.

As a simple example, we consider a quantum liquid of so-called
Fibonacci anyons \cite{Bouwknegt99,Read_Rezayi99,Slingerland01}. 
Here there are only two possible labelings, 
namely the
trivial particle ${\bf 1}$ and the Fibonacci anyon $\tau$. 
At any trivalent vertex of the skeleton lattice, there is a constraint forbidding the appearance
of only a single $\tau$-anyon on the three edges connected to the vertex,
allowing the following possibilities:
\begin{figure}[h]
\includegraphics[width=0.4\columnwidth]{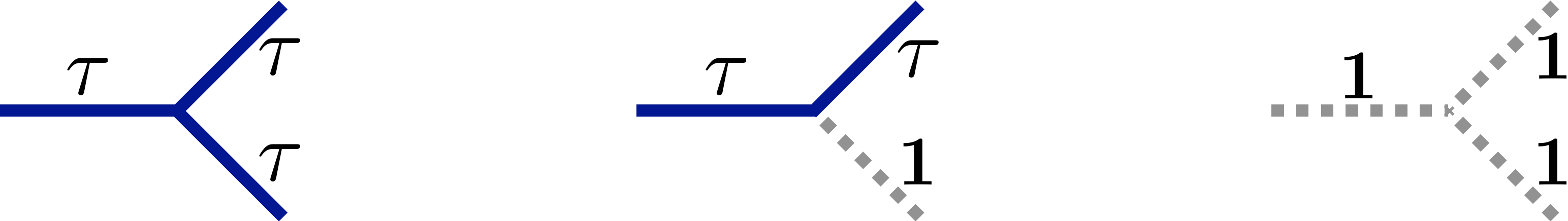}
\end{figure}

\noindent
Due to this constraint the edges occupied by a $\tau$-anyon form a closed, trivalent net 
known as a `string net' \cite{LevinWen05}.
One might as well identify the two degrees of freedom (${\bf 1},\tau$) with the two states of a 
spin-1/2 ($\uparrow,\downarrow$) and 
thus the same states
can be viewed as representing the ground states of a 
Hamiltonian with three-spin interactions enforcing the vertex constraint above 
(no single $\downarrow$-spin around a vertex)
\cite{Zoller}.

Returning to our model, we can now specify its microscopic terms
\begin{equation}
  H = - J_e \sum_{{\rm edges} \ e} \delta_{\ell(e),{\bf 1}} - J_p \sum_{{\rm plaquettes} \ p} \delta_{\phi(p),{\bf 1}} \,.
  \label{eq:Hamiltonian}
\end{equation}
The first term favors a trivial label $\ell(e)={\bf 1}$ on the edge $e$ corresponding to the no-flux state.
The second term favors the no-flux state $\phi(p) = {\bf 1}$ for the plaquette $p$. 
When expressed in terms
of the labels $\ell(e)$, the plaquette flux $\phi(p)$ is a complicated, but local expression involving the 
twelve edges connected to the vertices surrounding a plaquette, see Fig.~\ref{Fig:Model}, 
and is explicitly given in the supplementary material.
In the absence of the first term ($J_e=0$)  the plaquette term will effectively close all holes, and the 
ground state of the above Hamiltonian describes that of the quantum liquid on two parallel sheets as 
illustrated in Fig.~\ref{Fig:Transition2D}. 
The latter is precisely the string-net model first introduced by Levin and Wen \cite{LevinWen05},
which is also closely related to another model of string nets discussed recently by Fendley
\cite{Fendley}.
Similarly, in the absence of the plaquette term, $J_p=0$, the edge term with coupling constant $J_e$ 
will close off  all the \lq tubes\rq \ thus leading
to the ground state of the quantum liquid on multiple disconnected spheres as 
illustrated in Fig.~\ref{Fig:Transition2D}. 
This edge term acts as a string tension in the 
string net model, or as a magnetic field in 
its spin model representation.

\begin{figure}[b]
\includegraphics[width=0.6\columnwidth]{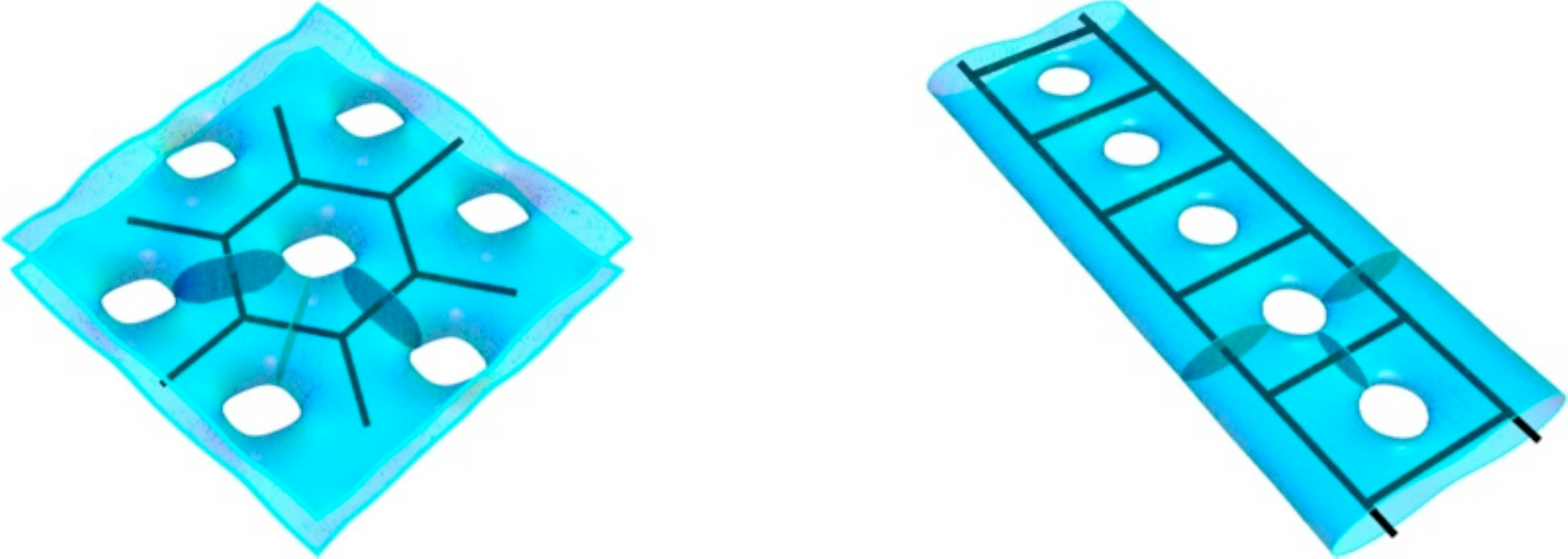}
\caption{{\it Microscopic model.} Our microscopic model energetically favors the flux-free states for the holes and tubes 
                (shaded) of the illustrated two-dimensional surfaces.
                For the surface with a triangular arrangement of holes shown in the left panel 
                the anyonic degrees of freedom in our model are associated with the edges 
                of the honeycomb lattice skeleton that surrounds the holes in the interior of the surface.
                For the linear geometry of holes on the right the skeleton lattice forms a ladder geometry.
                }
\label{Fig:Model}
\end{figure}

In the presence of both terms in the Hamiltonian, quantum fluctuations are introduced
which correspond to fluctuations of the surface. These fluctuations are
virtual processes where plaquettes or tubes close off and open depending on the flux
through them. We can visualize these fluctuations as local changes to the genus of the surface.
If the two terms in the Hamiltonian become comparable in strength, the competition between the two
drives a quantum phase transition between the two extremal 
topologies (see Fig.~1).  At this quantum phase transition the fluctuations of the surface become
critical and {\em the topology of the surface fluctuates} on all length scales. 
We can visualize the (imaginary) time-evolution of this quantum critical state as a `foam' in space-time,
which is reminiscent of the notion of a  quantum foam introduced by John Wheeler for fluctuations 
of 3+1 dimensional Minkowski space at the Planck scale \cite{Wheeler55,Wheeler57}. 

\begin{figure}[t]
\includegraphics[width=0.9\columnwidth]{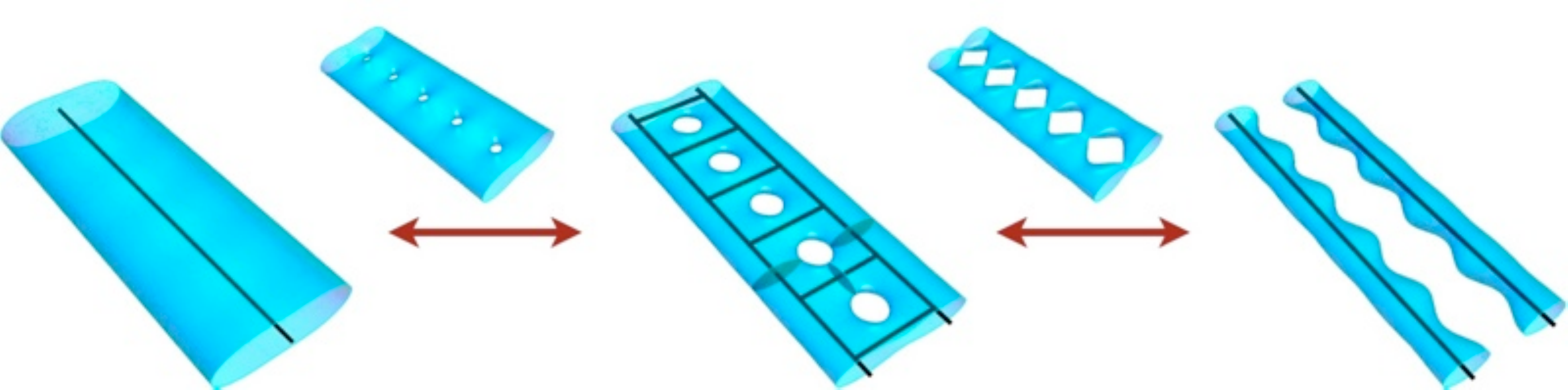}
\caption{{\it Phase transition in one dimension.} Illustration of the quantum phase transition driven by fluctuations 
	      of the surface topology in a linear geometry which connects the
	      extreme limits of a `single cylinder' (on the left) and `two cylinders'
	      (on the right).}
\label{Fig:Transition1D}
\end{figure}

To understand the nature of this transition, we first
focus on the linear geometry shown in Fig.~\ref{Fig:Model}.
In this geometry the Hamiltonian becomes
\begin{equation}
  H = - J_r \sum_{{\rm rungs} \ r} \delta_{\ell(r),{\bf 1}} - J_p \sum_{{\rm plaquettes} \ p} \delta_{\phi(p),{\bf 1}} \,,
  \label{eq:LadderHamiltonian}
\end{equation}
where the first term now only acts on the rungs between the holes 
(i.e. on those edges of the skeleton which separate two neighboring plaquettes),
in analogy to the original model.
This model exhibits a {\em continuous} quantum phase transition 
between the two extreme topologies shown in Fig.~\ref{Fig:Transition1D}. 
This continuous transition is driven by fluctuations of topology. 
It turns out that the gapless theory describing this transition can be {\em solved exactly}
as discussed in more detail below and explicitly in the supplementary
material.

The two extreme topologies connected by this transition in the linear geometry are as follows: 
In the limit of a vanishing rung term,
 $J_r=0$, the ground state is that of an anyonic quantum 
liquid  on a single cylinder where all the plaquettes are closed,
as shown on the left in Fig.~\ref{Fig:Transition1D}.
For Fibonacci anyons this ground state is two-fold degenerate, with either a $\tau$-flux or 
no flux through the cylinder. 
In the opposite limit of vanishing plaquette term, $J_p=0$, we can close off all the rungs and 
the ladder splits into two separate cylinders with a four-fold ground state degeneracy 
(either a $\tau$-flux or no flux in either of the cylinders),
as shown on the right in Fig.~\ref{Fig:Transition1D}.

\begin{figure}[b]
\includegraphics[width=0.75\columnwidth]{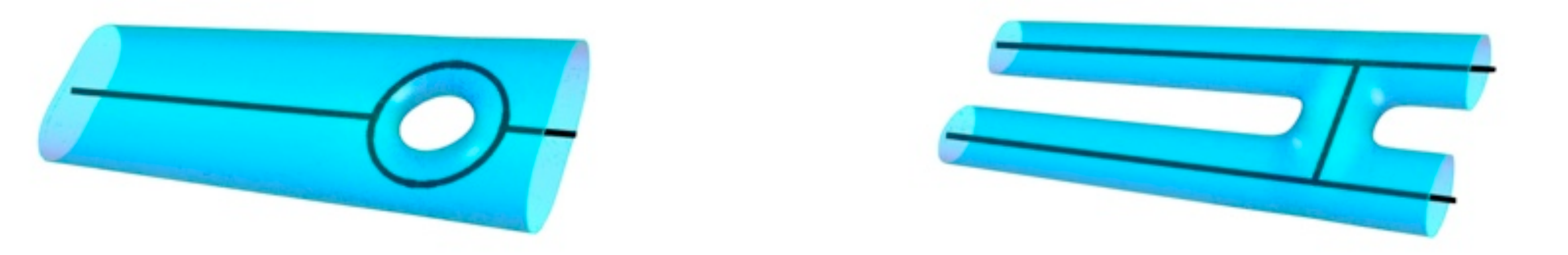}
\caption{{\it Excitations.} Plaquette (left) and rung (right) excitations above the two extreme ground states 
                illustrated in Fig.~\ref{Fig:Transition1D}.}
\label{Fig:Excitations}
\end{figure}

\begin{figure}[t]
\includegraphics[width=0.5\columnwidth]{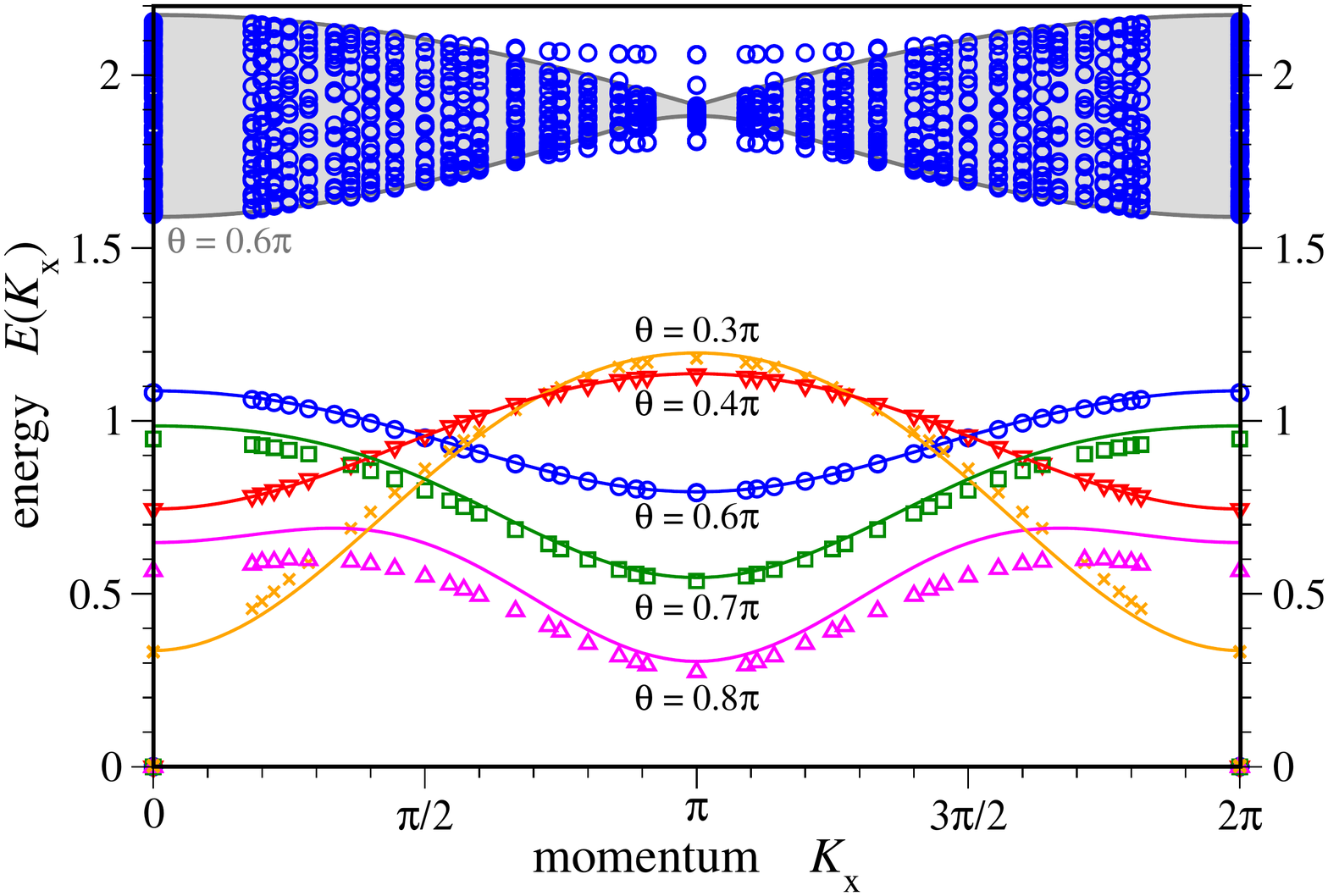}
\hspace{8mm}
\includegraphics[width=0.38\columnwidth]{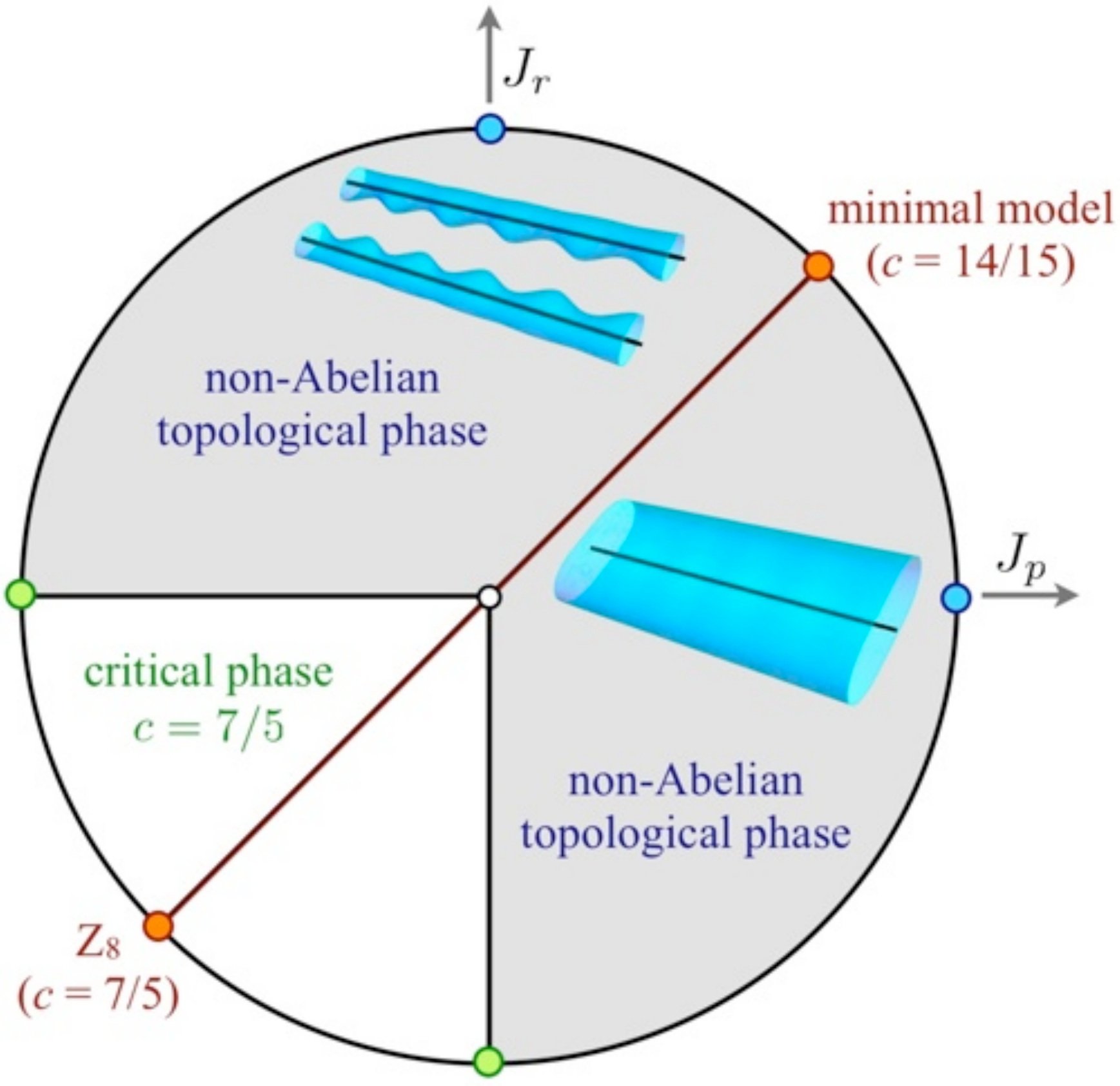}
\caption{{\it Left: Energy spectra.} Energy spectra of our microscopic model near the decoupling point ($\theta=\pi/2$).
                The rung excitations shown in Fig.~\ref{Fig:Excitations} form a gapped quasiparticle 
                band well below a continuum of states (shaded). 
                Open symbols show results from exact diagonalization of systems with 24 to 36 anyons.
                These bands are well described by second order perturbation 
                theory around the decoupling point shown as solid lines.
                {\it Right: Phase diagram.} The phase diagram of our microscopic model (\ref{eq:LadderHamiltonian}) where the
                couplings are parametrized as $J_p=\cos\theta$ and $J_r=\sin\theta$.
                The gapped topological phases are indicated by the shaded regions.
                The topology driven quantum phase transition occurs at the exactly solvable 
                critical point $\theta=\pi/4$.
                An extended critical phase is found in the region $\theta \in (\pi, 3\pi/2)$ around the 
                second solvable (critical) point $\theta=5\pi/4$. }
\label{Fig:Spectrum}
\end{figure}

In both limits excitations above these ground states are gapped quasiparticles with a gap 
of 
$J_p$ or $J_r$, respectively. 
The first excited state above the `single cylinder' ground state 
is a $\tau$-flux threading a single plaquette, which prevents it 
from being
closed as illustrated in 
Fig.~\ref{Fig:Excitations}a). 
In the opposite limit of the `two cylinder' ground state the first excited state is a $\tau$-flux through one
of the rungs, 
leaving this rung connecting the two cylinders as shown in Fig.~\ref{Fig:Excitations}b). Turning on a small coupling $J_r \neq 0$, or $J_p \neq 0$ respectively, these excitations delocalize, but remain gapped and form bands in the energy spectrum, as explicitly displayed  in Fig.~\ref{Fig:Spectrum}a). 
For large couplings, some of these excitations proliferate and their gap vanishes at the quantum phase transition mentioned above.

The full phase diagram 
is shown in Fig.~\ref{Fig:Spectrum}b), where we parameterize the two couplings on a circle 
as $J_p=\cos\theta$ and $J_r=\sin\theta$. 
Positive (negative) coupling constants indicate that the no-flux ($\tau$-flux)
states are energetically favored and the two extreme limits discussed above then correspond to the 
points $\theta=0$ and $\theta=\pi/2$ on the circle. 
The continuous phase transition between these two distinct topologies occurs for equal 
positive coupling strengths $J_r = J_p$, which corresponds to the point $\theta=\pi/4$ on the circle. 

We can visualize this critical point as a quasi one-dimensional quantum foam,
with topology fluctuations of the surface on all length scales.
As a first step, we have performed a detailed numerical analysis of this critical point 
using exact diagonalization of systems with up to 36 anyons.
The continuous nature of the phase transition reveals itself in a linear energy-momentum 
dispersion relation, which is indicative of conformal invariance.
A detailed analysis of the energy spectrum further allows to uniquely identify the 
corresponding conformal field theory (CFT), which in this case turns out to be the
$7$th member \cite{FootnoteModInv} of the famous series of so-called unitary
minimal CFTs \cite{FriedanQuiShenker} with central charge $c=14/15$.
This particular identification of a  conformal field theory is part of a broader scheme 
which connects the gapless theory of the topology driven phase transition with the nature 
of the underlying anyonic liquid.
In the present case of a quantum liquid of Fibonacci anyons we can make an
explicit connection between the (total) quantum dimension of the anyonic liquid 
and the central charge of the conformal field theory.

In fact, the Hamiltonian at this point is even exactly solvable. 
The key insight leading to this exact, analytical solution is the observation that the 
Hamiltonian of our topological model can be mapped precisely onto a particular 
version of the restricted-solid-on-solid (RSOS) model,   
which is exactly integrable and directly leads to the above-mentioned CFT
\cite{PasquierDAPartitionFunction}.
This mapping explicitly connects the Hamiltonian at this critical point with
an integrable Hamiltonian defined by the Dynkin diagram $D_6$
\begin{equation}
\includegraphics[width=0.3\columnwidth]{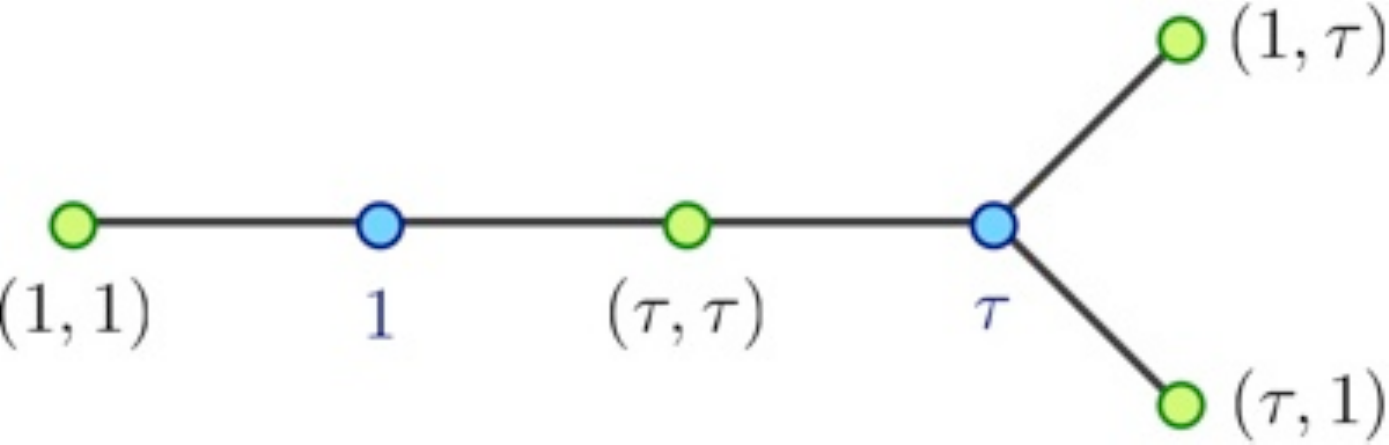} \quad.
\label{eq:dynkin}
\end{equation}
Here the particular labeling of the Dynkin diagram arises from the underlying
topological structure of our model. Specifically the labels describe the topological fluxes 
in the two extreme limits of the model as illustrated in Fig.~\ref{Fig:Transition1D}, 
with the limit of a `single cylinder' in the picture on the left corresponding to the 
blue circles in the Dynkin diagram and the limit of the `two cylinders' pictured 
on the right corresponding to the green circles.
This underlying structure also gives rise \cite{PasquierNPB1987}
to a representation of the Temperley-Lieb algebra  \cite{TemperleyLieb} 
which is characterized by the total quantum dimension $d=\sqrt{2 + \phi}$
of the anyonic liquid, where $\phi=(1+\sqrt{5})/2$ is the golden ratio.
A more detailed discussion of the exact solution is given in the methods section 
and the supplementary material.

Varying the couplings in our Hamiltonian there is another way of connecting the two
phases depicted in Fig.~\ref{Fig:Transition1D}, which is to change the sign of both
couplings in the Hamiltonian. For opposite sign the two terms now favor $\tau$-fluxes
through rungs and plaquettes, respectively, which again leads to a competition. 
Interestingly, we find that this competition results in an {\em extended}, critical phase 
separating the two topologically distinct phases, as depicted in the phase diagram of 
Fig.~\ref{Fig:Spectrum}.
For the full extent of this critical phase we again have topology fluctuations on all length
scales. However, the gapless theory describing this phase turns out to be in a different 
universality class as compared to the critical point discussed above.
These results can again be obtained through a combination of numerical and exact
analytical arguments, which are detailed in the supplementary material. 
In particular, there is another integrable point in this extended
critical phase for equal coupling strengths  $J_r = J_p$, which corresponds to the 
angle  $\theta=5\pi/4$ in the phase diagram of Fig.~\ref{Fig:Spectrum}, and is thus 
located exactly opposite of the one discussed above. Following a similar route one
can map the Hamiltonian at this second integrable point to another variant of the 
RSOS model associated with the Dynkin diagram $D_6$. 
The gapless theory at this point then turns out to be exactly the $Z_8$ parafermion 
CFT with central charge $c=7/5$.
The stability of this gapless theory away from the integrable point is due to an 
additional symmetry of our model \cite{GoldenChain,su2k}.
Numerically, we find that it extends throughout the whole region where both couplings 
favor the $\tau$-flux states all the way to the points $\theta=\pi$ and $\theta=3\pi/2$,
where there is no longer a competition of the two terms of the Hamiltonian
and the ground states have fluxes either through all plaquettes or rungs, respectively. 

Returning to our original discussion of the model (1) on the surface in Fig.~\ref{Fig:Transition2D}, 
the question arises  whether we can understand the nature of the quantum phase transition here as well. 
We can explicitly address this question in the context of another kind of anyons, the so-called semions
\cite{Semions}. 
Again, there are two possible labelings, the trivial particle ${\bf 1}$ and the semion $s$. The constraint now only allows zero or two semion particles $s$ at any trivalent vertex. 
The set of edges carrying a semion $s$ form loops instead of nets
and give rise to what is known as a `loop gas' \cite{Kitaev03,Freedman05}. 
In its spin-1/2 representation  (where $\uparrow,\downarrow$
now stand for ${\bf 1}$ and $s$)
this model is known as the honeycomb version of the toric code \cite{Kitaev03}, where the string tension $J_e$ corresponds to a magnetic field.
This model exhibits a continuous quantum phase transition in the 3D Ising universality class
\cite{Fradkin79,Trebst07} with topology fluctuations on all length scales. 
Mapping the 2+1 dimensional semion system to its three-dimensional classical counterpart, 
the quantum foam then corresponds to the critical fluctuations of domain walls in a 3D 
Ising model at its critical point.
For other kinds of anyonic liquids, the nature of the topology changing transition is in general unknown
and remains an intriguing open problem with the possibility of new universality classes. 
For a liquid of Fibonacci anyons there has been a recent discussion of quantum critical behavior
by Fendley from the perspective of ground-state wavefunctions and their respective correlators in terms of conformal field theory \cite{Fendley}. 

\begin{figure}[b]
\includegraphics[width=0.48\columnwidth]{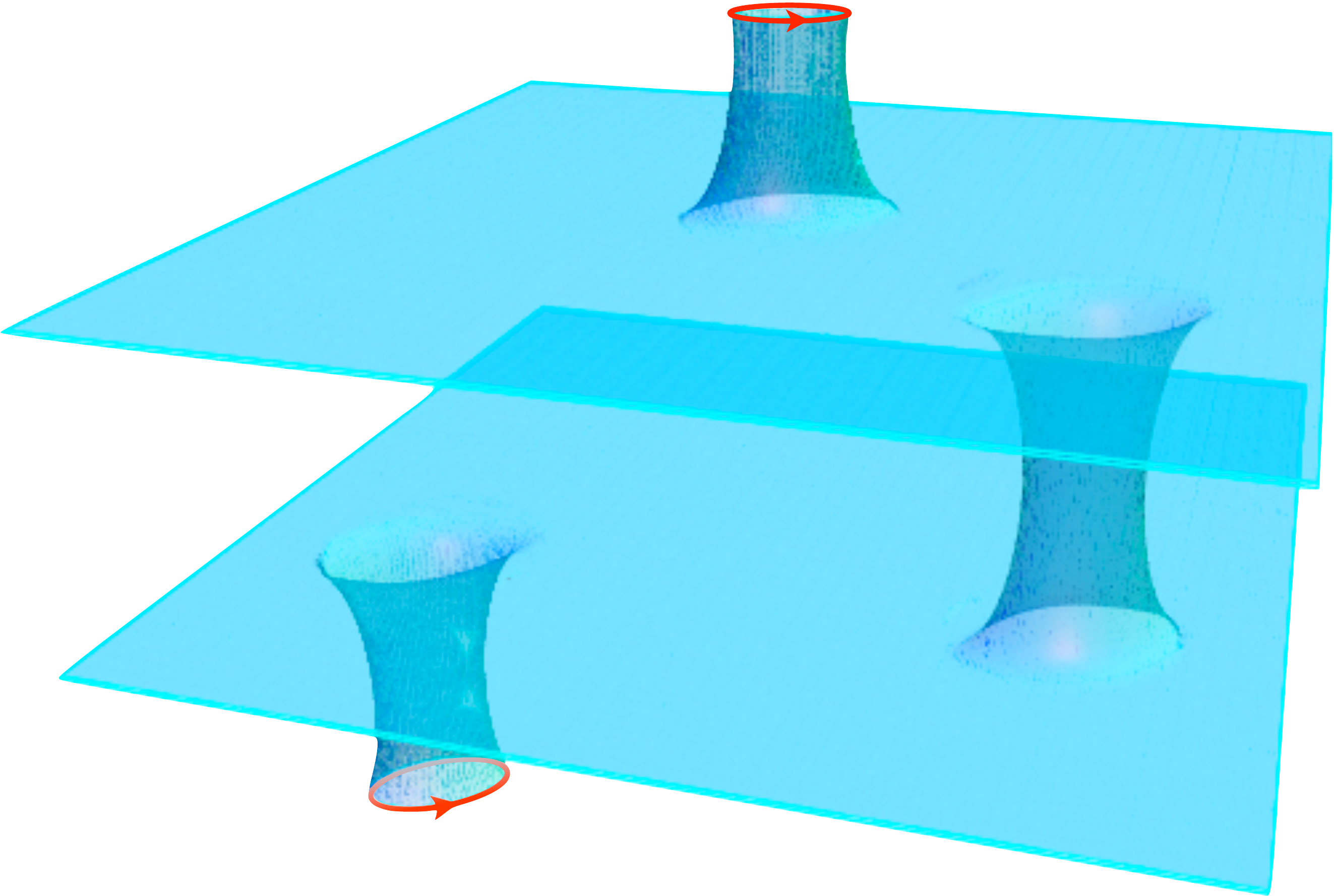}
\vskip -1mm
\caption{{\it Excitations of the anyonic liquid}. 
                Vortex excitations of the liquid indicated by the `chimneys' posses a chiral edge
                mode. 
               }
\label{Fig:ChiralExcitations}
\end{figure}

Finally, in order to explore the broader context of our models we complete our analysis
by considering the complete set of possible excitations present in these models.
An excitation different from the ones already discussed arises when relaxing the constraint which 
for every trivalent vertex of the skeleton lattice forbids the occurrence of a single $\tau$-flux.
If we allow for this possibility, we are left with a $\tau$-flux entering the vertex through one tube, 
but not leaving it through another tube in the skeleton plane as illustrated in Fig.~\ref{Fig:Model}.
Instead we can think of the remaining $\tau$-flux at such a vertex as leaving through one of the 
liquid sheets surrounding the skeleton lattice. 
This piercing of the liquid by a $\tau$-flux corresponds to a vortex excitation of the liquid and 
is illustrated as a `chimney' in Fig.~\ref{Fig:ChiralExcitations}. 
These vortex excitations break time-reversal symmetry and turn out to all possess the same
chirality (indicated by the red arrow in Fig.~\ref{Fig:ChiralExcitations}). 
This is only possible if the anyonic liquid on a given sheet itself possess a given chirality. 
Since the entire system exhibits time-reversal symmetry, this means that the two anyonic
liquids on the two sheets must have opposite chirality.
Vortices associated with chimneys on opposite sheets thus also have opposite chirality
as illustrated in Fig.~\ref{Fig:ChiralExcitations}. 
(In fact, a vortex in one sheet can be related to a vortex in the opposite sheet by dragging a vortex
through a `hole' connecting the two sheets. Moreover, we can create a `hole' connecting the sheets
by glueing together two vortex excitations on opposite sheets.)
This conceptual perspective of two anyonic liquids with opposite chirality giving rise to a 
time-reversal invariant model connects with and allows a visualization of a more abstract 
mathematical description of these models, 
namely doubled non-Abelian Chern-Simons theories \cite{Freedman04}.
It remains an intriguing question to formulate our topology driven phase transitions within
such a field theoretical framework.

In this manuscript we have developed a general, unifying framework to formulate topological aspects
of quantum states of matter for systems preserving time-reversal symmetry and of their phase
transitions. In a simple and  intuitive picture they are described in terms of fluctuations of two-dimensional surfaces and their topology changes.
Our framework gives a new perspective on how to broadly discuss quantum phase transitions out of 
topologically ordered states of matter, which has so far been largely unexplored territory due to
the lack of a local order parameter description amenable to a Landau-Ginzburg-Wilson theory.
This description of time-reversal invariant anyonic quantum liquids is also expected to advance our understanding of spin liquid states and their phase transitions in recently proposed materials of 
frustrated quantum magnetism and other strongly correlated systems. 
Our unifying perspective on string nets and quantum loop gases will also allow to construct 
a large variety of new microscopic models for topological phases.

{\em Acknowledgments.--}
We thank M. Freedman, X.-G. Wen, and P. Fendley for stimulating discussions. 
Our numerical simulations were based on the ALPS libraries \cite{ALPS}.  
A. W. W. L. was supported, in part, by NSF DMR-0706140.


\section*{\large Methods}

\noindent {\bf Identification of conformal field theories.} 
To characterize the conformal field theory (CFT) of the critical points in the linear (ladder) geometry, 
we rescale and match the finite-size energy spectra obtained numerically by exact diagonalization
for systems with up to $L=36$ anyons to the form of the  spectrum of a CFT,
\begin{equation}
E = E_1L +\frac{2\pi v}{L} \left (-\frac{c}{12}+ h+\bar{h} \right ),
\label{CFT_energy_levels}
\end{equation}
where the velocity $v$ is an overall scale factor, and $c$ is the central  charge of the CFT.
The scaling dimensions $h+\bar{h}$ take the form $h=h^0+n$, $\bar{h}=\bar{h}^0+\bar{n}$,
with $n$ and $\bar{n}$ non-negative integers, and $h^0$ and $\bar{h}^0$ are the holomorphic and antiholomorphic conformal weights of primary fields in a given CFT with central charge $c$. 
The momenta (in units $2\pi/L$) are such that $k_x=h-\bar{h}$ or $k_x=h-\bar{h}+L/2$.
Using this procedure we find that for the critical point at $\theta=\pi/4$ the rescaled energy spectrum matches the assignments (\ref{CFT_energy_levels}) of the $7$th member \cite{FootnoteModInv} of the famous series of so-called unitary minimal CFTs \cite{FriedanQuiShenker} with central charge $c=14/15$. 
Similarly, at the point $\theta=5\pi/4$ we find the rescaled energy spectrum to match
that of the $Z_8$ parafermion CFT with central charge $c=7/5$. 
For the calculated energy spectra and the details of these assignments we refer to the 
supplementary material.
\\

\noindent {\bf Exact analytical solution.} 
The Hamiltonian in Eq.~\eqref{eq:LadderHamiltonian} can be solved exactly 
for interaction strengths corresponding to angles  $\theta=\pi/4$ and $\theta=5\pi/4$ 
in the phase diagram of Fig. \ref{Fig:Spectrum}. This exact, analytical
solution of the gapless theories at these points unambiguously demonstrates the 
continuous nature of the related quantum phase transitions and points to generalizations
of these gapless theories for other kinds of anyonic liquids. 
The key observation underlying this exact solution
is the emergence of the $D_6$ Dynkin diagram from the topology of the surface 
associated with the ladder model as depicted in the right part of Fig.~\ref{Fig:Model}.
Each labeling of the edges of the `skeleton' graph  
which corresponds to that surface, denotes one of the states spanning the Hilbert space
of the system.  
A crucial step is to consider a different `pants decomposition'
\cite{MooreSeiberg}
of this surface
and to perform a  {\em basis change} to a new basis which corresponds to the labeling of the 
skeleton lattice of this alternative pants decomposition.
Explicitly, this basis transformation can be written as
\begin{figure}[h]
\begin{center}
  \scalebox{0.42}[0.42]{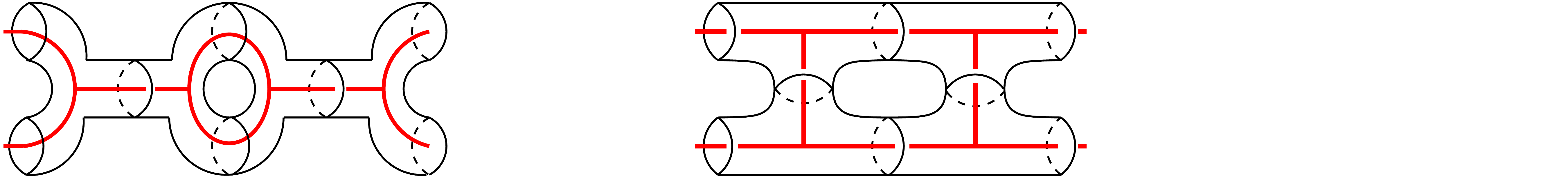} 
\end{center}
\end{figure}

\noindent
Here, $({F^a_{b c d}})_{a'}^{b'} $ denotes the so-called $F$-matrix, which is a generalization
of the familiar $6j$ symbols of angular momentum coupling in conventional quantum 
mechanics and is 
known for any anyonic liquid 
\cite{QuantumSixJ-Reshetikhin}.
Note that associated with the even-numbered indices of these labels, which correspond to the 
original rung labels $c_i$ on the right, there is the flux through the cross-section of the surface
on the left, denoted by a label $d_i=1$ or $d_i=\tau$. 
Similarly, associated with the odd-numbered indices, which correspond to the original plaquettes, 
there is a {\it pair} of fluxes through the two cross-sections of the surface 
at the position of the plaquette on the left, denoted by a {\em pair of labels}, $(a_i,b_i)$.
This pair of labels can assume four values, i.e., $(a_i,b_i)= \{ (1,1); (\tau,1); (1,\tau); (\tau,\tau) \}$.
The (fusion) constraints at the vertices where the labels $(a_i,b_i)$ and $d_{i\pm 1}$ 
meet then turn out to be precisely the condition
that they be adjacent nodes on the $D_6$  Dynkin diagram of Eq.~\eqref{eq:dynkin}.
For example, a local label $(a_i,b_i)=(\tau,\tau)$
at an odd-numbered index $i$  only allows for labels $d_{i-1}=1$ and $d_{i-1}=\tau$
at the neighboring even-numbered indices.
This is reflected in the Dynkin diagram by the appearance of a line  that connects the label $(\tau,\tau)$ to both labels $1$ and $\tau$.
The importance of the just described basis change consists in the fact that in the new basis
the rung and plaquette terms,
$ H^r_{i}$ and $H^p_{i}$, respectively,
of our ladder Hamiltonian
 \begin{align}
\label{Ham2}
H &= -
J_r
\sum_{i\;{\rm even}}
  H^r_{i}  -
J_p
\sum_{i\;{\rm odd}}
 H^p_{i}
\; ,
 \end{align}
turn out to
have precisely the form of a known 
representation~\cite{PasquierNPB1987} 
of the Temperley-Lieb algebra~\cite{TemperleyLieb}
associated with the $D_6$ Dynkin diagram,
\begin{equation}
{\bf e}_i^2=D \  {\bf e}_i \,,
\quad
{\bf e}_i  {\bf e}_{i\pm1} {\bf  e}_i = {\bf  e}_i \,,
\qquad 
[{\bf e}_i, {\bf e}_j]=0 \quad {\rm for} \ |i-j|\geq 2 \,,
\label{TemperleyLiebRelations}
\end{equation}
where
\begin{equation}
\label{HRandHP}
{\bf e}_i  = 
\begin{cases}
\ \ 
D \  H^r_{i}  \quad \quad  {\rm for}\  i \ {\rm   even},
\\ 
\ \
D \ H^p_{i}  \quad \quad  {\rm for}\  i \ {\rm   odd}.
\end{cases}
\end{equation}
The characteristic `D-isotopy' parameter of this Temperley-Lieb algebra, 
$D=\sqrt{1+\varphi^2} = 2 \cos (\pi/10)$,
is precisely the total quantum dimension of the underlying Fibonacci anyon liquid.
We have thereby established a remarkable, explicit connection of the one parameter of this emerging algebraic structure, the `D-isotopy' parameter of this Temperley-Lieb algebra, 
and the single most characteristic parameter of the underlying anyonic liquid, 
namely its total quantum dimension. This observation points to a generalization
of such a connection for other quantum liquids.
Written in this form, the resulting Hamiltonian for the Fibonacci anyon liquid 
turns out to be precisely that of the
(integrable) restricted solid-on-solid (RSOS) statistical mechanics
lattice model based on the $D_6$-Dynkin diagram~\cite{PasquierNPB1987},
as obtained in the standard fashion
from the transfer matrix of the RSOS lattice model. 
For further details we refer to the supplementary material.
 \\

\noindent {\bf General framework.} 
We have explicitly formulated the concept of a topology driven phase 
quantum phase transition mainly in the context of a single anyon theory, 
namely that of Fibonacci anyons.
However, we note that these concepts apply in great generality to any anyon
theory in which there can be an arbitrary number of anyons subject to a set of
fusion rules / constraints.
\\

\newpage

\begin{center}
{\large  {\bf Supplementary Material}}
\end{center}

\section{Fibonacci anyons}
\label{SectionFibonacciAnyons}
   
\noindent
The degrees of freedom in our microscopic models are so-called Fibonacci anyons, 
one of the simplest types of 
non-abelian anyons \cite{Read_Rezayi99,preskill}. 
The Fibonacci theory
has two distinct particles, the trivial state $1$ and the  Fibonacci anyon 
$\tau$, which can be thought of as
a generalization (or more precisely a `truncated version')
of an  `angular momentum'  when viewing the 
Fibonacci theory as a certain 
deformation~\cite{deformation} of SU(2).
We will now make this notion more precise
and illustrate it in detail.
In analogy to the ordinary angular momentum coupling rules, we can write down a set
of `fusion rules' for the anyonic degrees of freedom
which are analogs of the Clebsch-Gordon rules for
coupling of ordinary  angular momenta,
\begin{equation}
 1 \times 1 = 1 \quad\quad\quad
 1 \times \tau = \tau = \tau \times 1 \quad \quad \quad
  \tau \times \tau  =1+\tau \,,
 \label{fusion_rules} 
\end{equation}
where the last fusion rule reveals 
what is known as 
the non-abelian character of the Fibonacci anyon:
Two Fibonacci anyons $\tau$ can fuse to either the trivial particle or to another 
Fibonacci anyon.
In more mathematical terms, these fusion rules can also be expressed 
 by means of so-called
fusion matrices $N_j$ whose entries  $(N_j)^{j_1}_{j_2}$ 
 equal to one if
and only if
the fusion of anyons of types $j_1$ and $j_2$ into $j$ is possible.
The fusion rules are related to the 
so-called
`quantum dimensions' $d_j$  of the anyonic
 particles  by
 \begin{equation}
   N_j |d_j\rangle = d_j |d_j\rangle \,, 
   \label{fusion_matrices}
 \end{equation}
 where  $|d_j\rangle$ is 
the (`Perron Frobenius')
eigenvector corresponding to the largest
positive 
eigenvalue of the $2\times 2$-matrix $N_j$.
[The sense in which these numbers are `dimensions'
will become  apparent
in section
\ref{subsectionExplicitExpression}
below.]
For the particles in the Fibonacci theory the quantum dimensions are $d_1=1$ and 
$d_{\tau} = \varphi \equiv (1+\sqrt{5})/2$ and
the total quantum dimension of the theory is then given by 
$D= (\sum_j  d_j^2)^{1/2}=\sqrt{1+\varphi^2}$.

 To define our Hamiltonian, some additional indegredients of the theory of anyons are required.
In analogy to the $6j$-symbols for ordinary SU(2) spins, there exists a basis transformation $F$ 
that  relates the two differents ways three anyons can fuse to a fourth anyon, depicted as
\begin{equation}
\parbox{2cm}{ \scalebox{0.3}[0.3]{
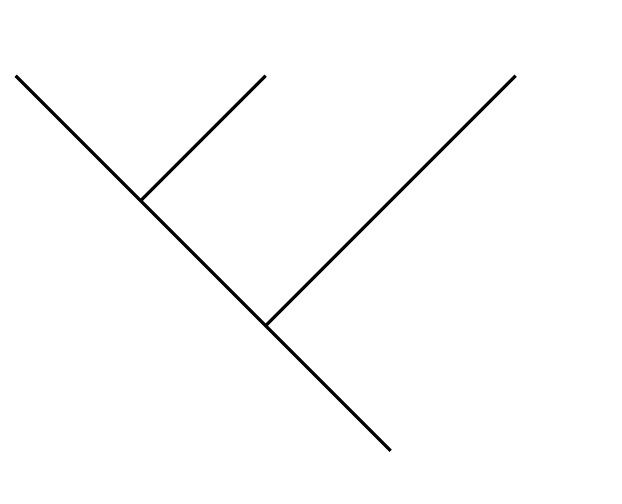}}
 = \sum_f (F_{abc}^d)_e^f  \parbox{2cm}{ \scalebox{0.3}[0.3]{
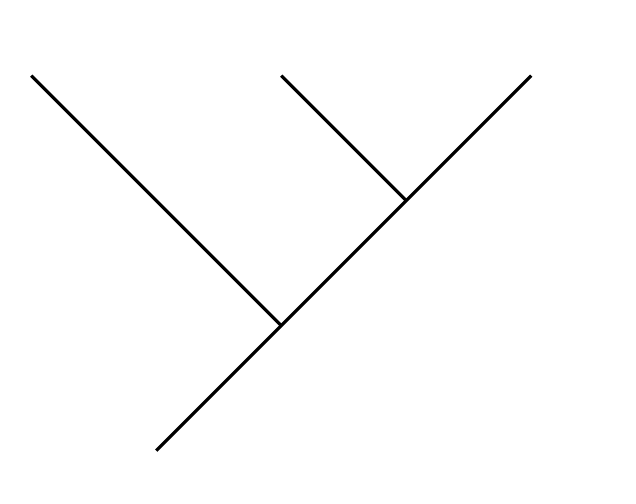}}.
\label{Fmatrix_eq}
\end{equation}
The left hand side (l.h.s.)  represents the quantum state that
arises when anyon $a$ first fuses with anyon $b$ into
an anyon of type $e$, which, subsequently,
fuses with anyon $c$ into an anyon of type $d$.
Similarly, the right hand side (r.h.s.) denotes the quantum states
that arises when anyon $b$ first fuses with anyon $c$
into anyon type $f$ which, in turn,
fuses with anyon $a$ into anyon type $d$. Whilst keeping
all external labels, the types of the three anyons ($a,b,c$)
as well as the resulting anyon type $d$ fixed, 
the states on the l.h.s and r.h.s. are 
fully specified by the
labels $e$ and $f$, respectively.
Eq. (\ref{Fmatrix_eq}) says the so-specified states are linearly related to each
other by the so-called 
$F$-matrix \cite{RefFMatrix}
with matrix elements $ (F_{abc}^d)_e^f$.

In general, the $F$-matrix is uniquely defined
(up to \lq gauge transformations\rq)
 by the fusion rules through a consistency
relation called the pentagon equation \cite{MooreSeiberg}. 
Similarly, the braiding properties of anyons are given
by the so-called $R$-matrix (which however is not needed here) that is uniquely determined
by the hexagon equation \cite{MooreSeiberg}.

For the Fibonacci theory, it is straightforward to verify that in most cases there is only  
one term on the right-hand-side in Eq.~(\ref{Fmatrix_eq}), e.g. by choosing two or three 
out of the four anyons $a$, $b$, $c$, $d$ to be $\tau$-anyons.
For these cases the consistency with the pentagon and hexagon relations then yields 
that the corresponding $F$-matrix elements equal to $1$.
There is only one configuration that gives rise to $F$-matrix elements that are non-trivial:
If all anyons are $\tau$-anyons, e.g. $a=b=c=d=\tau$, both the $1$- and the $\tau$-fusion 
channels appear, and the $F$-matrix takes the explicit form
\begin{equation}
  F_{\tau \tau \tau}^{\tau} = \left (\begin{array}{cc}
(F_{\tau \tau \tau}^{\tau})_1^1&(F_{\tau \tau \tau}^{\tau})_{\tau}^1\\ (F_{\tau \tau \tau}^{\tau})_1^{\tau}&(F_{\tau \tau \tau}^{\tau})_{\tau}^{\tau}
   \end{array} \right ) = \left (\begin{array}{cc}
\varphi^{-1}&\varphi^{-1/2}\\ \varphi^{-1/2}&-\varphi^{-1}
   \end{array} \right ).
   \label{F_matrix_Fib}
 \end{equation}
As a final ingredient to explicitly derive our Hamiltonian, we have to introduce the
so-called modular
$S$-matrix 
that relates the anyon ``flux" of species $b$ through an anyon loop of species $a$
 to the case without anyon loop by
 \begin{equation}
 \parbox{1.5cm}{\scalebox{0.3}[0.3]{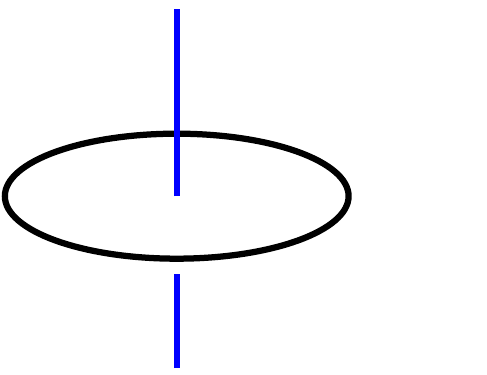}} =\;\; \frac{S_a^b}{S_1^b} \hspace{3mm}
  \parbox{1.8cm}{\scalebox{0.3}[0.3]{\begin{picture}(0,0)%
\includegraphics{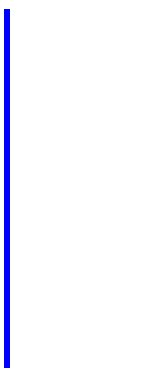}%
\end{picture}%
\setlength{\unitlength}{3947sp}%
\begingroup\makeatletter\ifx\SetFigFont\undefined%
\gdef\SetFigFont#1#2#3#4#5{%
  \reset@font\fontsize{#1}{#2pt}%
  \fontfamily{#3}\fontseries{#4}\fontshape{#5}%
  \selectfont}%
\fi\endgroup%
\begin{picture}(707,1791)(1168,-994)
\put(1276,464){\makebox(0,0)[lb]{\smash{{\SetFigFont{25}{30.0}{\familydefault}{\mddefault}{\updefault}{\color[rgb]{0,0,1}$b$}%
}}}}
\end{picture}%
}} \,.
 \label{S_eq}
 \end{equation}
For the case of  Fibonacci anyons, the $S$-matrix takes the explicit form
\begin{equation}
S =  \left ( \begin{array}{cc} S_1^1&S_{\tau}^1\\ S_1^{\tau} &S_{\tau}^{\tau}\end{array}\right )= 
\frac{1}{D} \left ( \begin{array}{cc} 1&\varphi\\ \varphi &-1\end{array}\right ).
\label{S_matrix_Fib}
\end{equation}
There is an important relationship between the modular $S$-matrix and the matrix
encoding the fusion rules, introduced in the paragraph above Eq. (\ref{fusion_matrices}):
the modular $S$-matrix {\it diagonalizes} the fusion rules,
the \lq Verlinde Formula\rq,
\begin{equation}
S^b_{b'}
\ (N_a)^{b'}_{c'} \ {S^\dagger}^{c'}_c 
\ \ = \ \ 
\delta^b_c \ \ {S^a_b \over S^1_b} \,,
\label{Verlinde}
\end{equation}
(repeated indices are summed)
where $S^\dagger$ denotes the adjoint of the unitary matrix $S$.
The eigenvalues of the matrix $(N_a)$ are thus ${S^a_b \over S^1_b}$,
and the largest (positive) eigenvalue,
the quantum dimension $d_a$,
can be seen to be
\begin{equation}
d_a = {S^a_1 \over S^1_1} \,.
\label{QuantumDimensionSMatrix}
\end{equation}
Due to the unitarity of the modular $S$-matrix one immediately
checks
that the total quantum dimension equals 
\begin{equation}
D ={1 \over S^1_1}.
\label{TotalQuantumDimensionSMatrix}
\end{equation}

\section{The Ladder Model}

In this section we will discuss details of the ``ladder model'' in a one-dimensional geometry, 
whose Hamiltonian  is given by Eq.~(2) in the main part of
the  paper. We start by defining the Hamiltonian in detail, and then discuss the gapped
topological phases, critical phases, and the exact solutions.

\subsection{The Hamiltonian}

\subsubsection{Explicit expression}
\label{subsectionExplicitExpression}

\begin{figure}[t]
\begin{center}
\scalebox{0.4}[0.4]{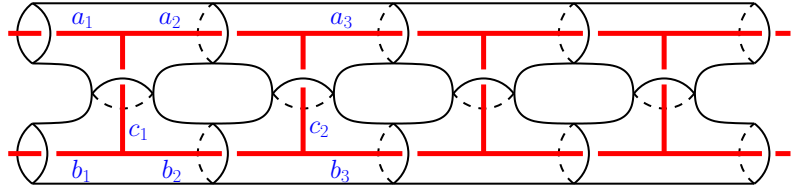}
\caption{The high-genus surface in a ladder geometry and the `skeleton' of the fusion graph 
                that defines the Hilbert space. The trivial particle $1$ or the Fibonacci anyon $\tau$ 
                can occupy the links of the ladder skeleton, subject to the vertex constraints given by 
                the fusion rules of Fibonacci anyons.}
\label{ladder_surface}
\end{center}
\end{figure}

To establish a notation for the basis states we consider the 
skeleton lattice inside the high-genus ladder geometry as shown in Fig.~\ref{ladder_surface}.
The basis states are given by all admissible labeling of the edges of the skeleton with $1$ or $\tau$ particles, subject to the vertex constraints given by the fusion rules. The number of basis states, $B_L$, of the ladder with $L$ plaquettes and  periodic boundary conditions  is given by
\begin{equation}
B_L = \sum_{\{a_i,b_i,c_i \} }  (N_{c_1})_{a_1}^{a_2} (N_{c_2})_{a_2}^{a_3} \ldots  (N_{c_L})_{a_L}^{a_1}
(N_{c_1})_{b_1}^{b_2} (N_{c_2})_{b_2}^{b_3} \ldots  (N_{c_L})_{b_L}^{b_1}
= \sum_{\{i_1,\ldots i_L\}} [{\rm Tr} (N_{i_1} N_{i_2} \ldots  N_{i_L})]^2 \,,
\label{dim_hilbert}\end{equation}
where $N_i$ are the fusion matrices of Fibonacci theory as introduced above. 
The largest eigenvalue of the matrix $N_i$ is the quantum dimension $d_i$.
Thus, the leading behavior of the traces for large $L$ is,
\begin{equation}
B_L 
\sim
\sum_{\{i_1\ldots i_L \}} (d_{i_1} d_{i_2} \ldots  d_{i_L})^2 =
\sum_{k=0}^L (d_1^2)^{L-k} (d_{\tau}^2)^k  = (1+\varphi^2)^L = D^{2L}.
\end{equation}
The Hilbert space thus grows asymptotically, for large $L$,
as a power of the square of the total quantum dimension $D^2$.

The Hamiltonian (as given in Eq.~(2) of the main part of the paper)
\begin{equation}
H = - J_r \sum_{{\rm rungs} \ r} \delta_{\ell(r),1} - J_p \sum_{{\rm plaq} \ p} \delta_{\phi(p),1} 
\label{Eq:Hamiltonian}
\end{equation}
consists of two non-commuting  terms, the rung term which is diagonal in the chosen basis, and the plaquette 
term which depends on the four edges of the plaquette $p$, and the four adjoining edges.  
By inserting an additional anyon loop of type $s$ into the center of the plaquette, we can project onto
the flux through this additional loop (and hence the flux through the plaquette) by
the following procedure (for a derivation see the following subsection)

\begin{equation}
\delta_{\phi(p),1} \left |
\parbox{2.3cm}{ \scalebox{0.37}[0.37]{
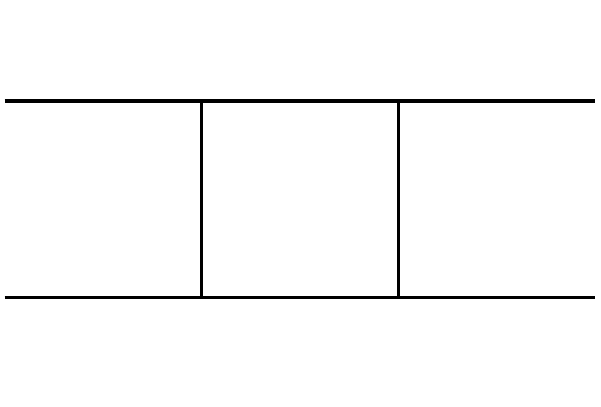 }}
\right \rangle
= \sum_{s=1,\tau} \frac{d_s}{D^2} \left |
\parbox{2.3cm}{ \scalebox{0.37}[0.37]{
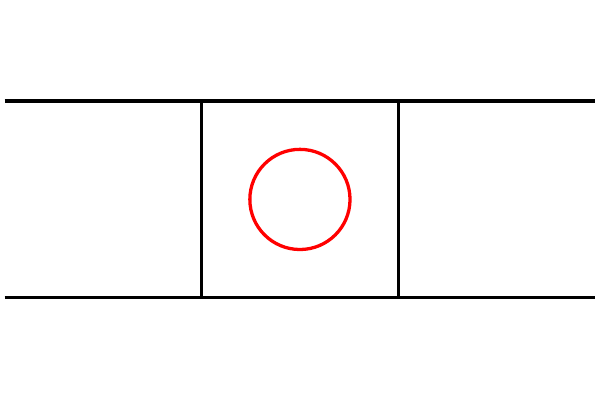 }}
\right \rangle.
\label{ProjectorImplemented}
 \end{equation}
 The additional $s$-loop is inserted by performing a sequence of $F$-transformations: 
 \begin{align}
& \left |
\parbox{2.3cm}{ \scalebox{0.37}[0.37]{
\input{plaq_term4.pdf_t} }}
\right \rangle = 
\sum_{\delta'}  (F_{\delta \delta s}^s)_1^{\delta'}  \left |
\parbox{2.3cm}{ \scalebox{0.37}[0.37]{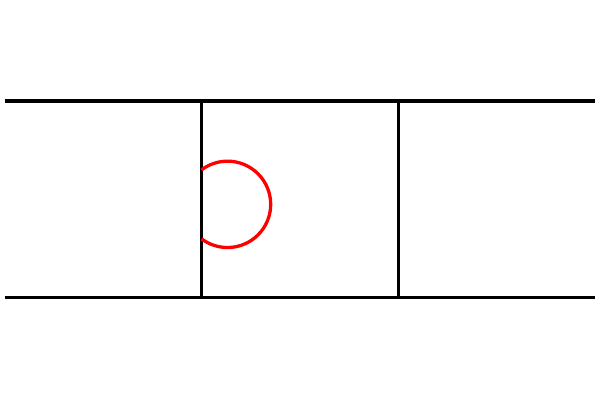 }}
\right \rangle
= \sum_{\delta',\gamma'}  (F_{\delta \delta s}^s)_1^{\delta'} (F_{d\delta' s}^{\gamma})_{\delta}^{\gamma'}
  \left |
\parbox{2.3cm}{ \scalebox{0.37}[0.37]{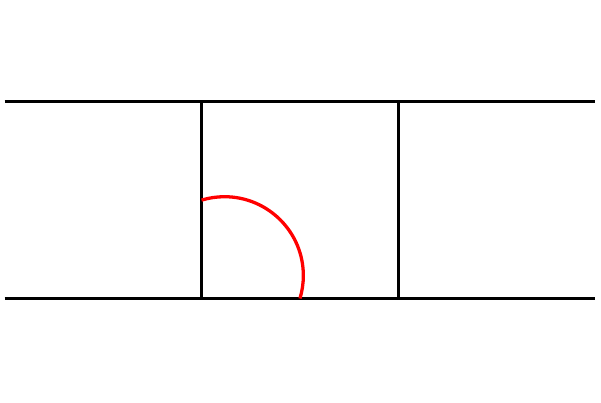 }}
\right \rangle
\\ \nonumber
 =&  \sum_{\delta',\gamma',\beta'}  (F_{\delta \delta s}^s)_1^{\delta'} (F_{d\delta' s}^{\gamma})_{\delta}^{\gamma'}
(F_{c\gamma's}^{\beta})_{\gamma}^{\beta'}
  \left |
\parbox{2.3cm}{ \scalebox{0.37}[0.37]{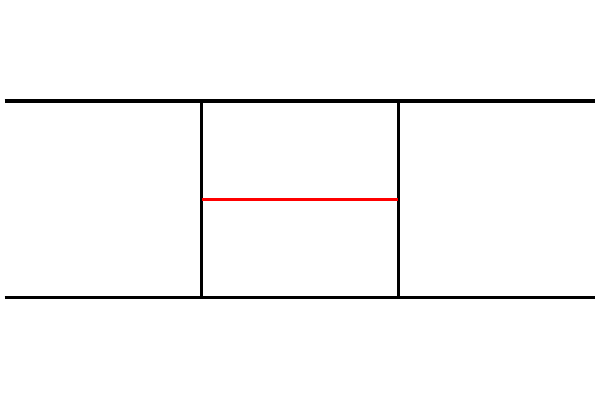 }}
\right \rangle
 =  \sum_{\delta',\gamma',\beta',\alpha'}  (F_{\delta \delta s}^s)_1^{\delta'} (F_{d\delta' s}^{\gamma})_{\delta}^{\gamma'}
(F_{c\gamma's}^{\beta})_{\gamma}^{\beta'} (F_{b\beta's}^{\alpha})_{\beta}^{\alpha'}
  \left |
\parbox{2.3cm}{ \scalebox{0.37}[0.37]{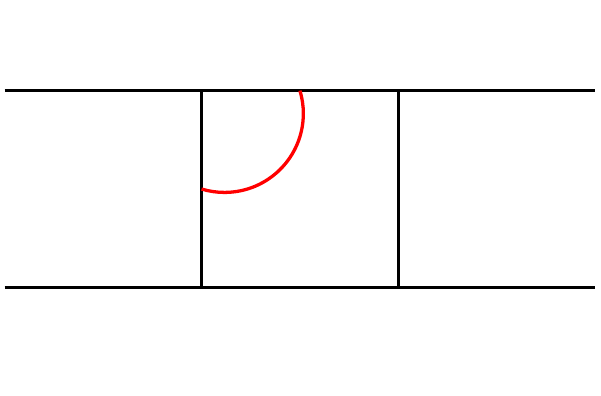 }}
\right \rangle  \\ \nonumber
=&  \sum_{\delta',\gamma',\beta',\alpha',m}  (F_{\delta \delta s}^s)_1^{\delta'} (F_{d\delta' s}^{\gamma})_{\delta}^{\gamma'}
(F_{c\gamma's}^{\beta})_{\gamma}^{\beta'} (F_{b\beta's}^{\alpha})_{\beta}^{\alpha'}
(F_{a \alpha' s}^{\delta})_{\alpha}^{m}
  \left |
\parbox{2.3cm}{ \scalebox{0.37}[0.37]{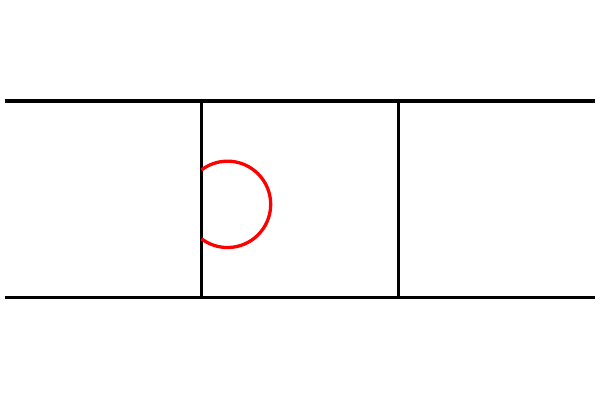 }}
\right \rangle. 
\end{align}
 Using the identities
 \begin{equation}
 \parbox{1.2cm}{ \scalebox{0.35}[0.35]{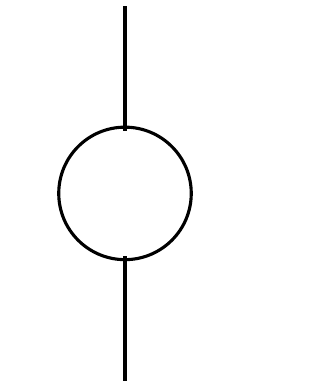 }}
 =\delta_{m \delta'}  \; (F_{\delta \delta s}^s)_1^{\delta'} 
  \hspace{0.7cm} \parbox{1cm}{ \scalebox{0.37}[0.37]{\begin{picture}(0,0)%
\includegraphics{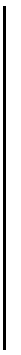}%
\end{picture}%
\setlength{\unitlength}{3947sp}%
\begingroup\makeatletter\ifx\SetFigFont\undefined%
\gdef\SetFigFont#1#2#3#4#5{%
  \reset@font\fontsize{#1}{#2pt}%
  \fontfamily{#3}\fontseries{#4}\fontshape{#5}%
  \selectfont}%
\fi\endgroup%
\begin{picture}(97,1694)(1179,-1133)
\put(1276,-286){\makebox(0,0)[lb]{\smash{{\SetFigFont{25}{30.0}{\familydefault}{\mddefault}{\updefault}{\color[rgb]{0,0,0}$\delta'$}%
}}}}
\end{picture}%
 }},
 \end{equation}
and $(F_{abc}^d)_e^f =(F_{bcd}^a)_f^e$,  we obtain the final expression 
 \begin{equation}
 \delta_{\phi(p),1} \left |
\parbox{2.3cm}{ \scalebox{0.37}[0.37]{
\input{plaq_term1.pdf_t} }}
\right \rangle
= \sum_{s=1,\tau} \frac{d_s}{D^2} 
\sum_{\alpha',\beta',\gamma',\delta'} 
 (F_{d\delta' s}^{\gamma})_{\delta}^{\gamma'}
(F_{c\gamma's}^{\beta})_{\gamma}^{\beta'} (F_{b\beta's}^{\alpha})_{\beta}^{\alpha'}
(F_{a \alpha' s}^{\delta})_{\alpha}^{\delta'}
\left  | \parbox{2.3cm}{ \scalebox{0.37}[0.37]{
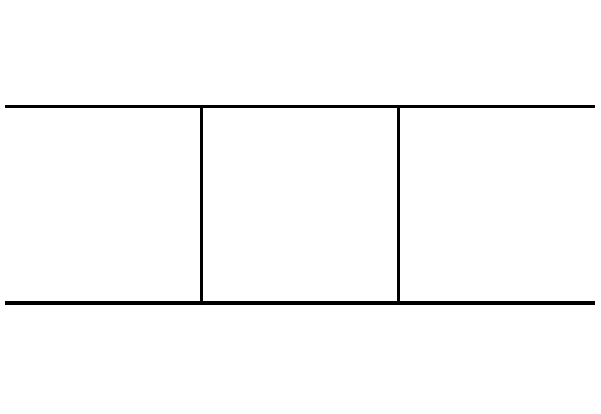 }}
\right \rangle. \label{plaq_term}
 \end{equation}

The ladder geometry has a local duality between the inside and outside: the inside of the rungs is dual  to the plaquettes. The only difference is that the rungs connect two different cylinders, while the plaquettes connect the same space (the ``outside''). The duality can be made exact by using ``twisted" boundary conditions where the ends of the ladder are
connected according to $a_1 =b_{L+1}$ and $b_1=a_{L+1}$
(so that the ladder looks like a Moebius strip).
Indeed, our exact diagonalization results confirm that the excitation spectra are identical 
under exchange of the couplings $J_r$ and $J_p$ for twisted boundary conditions. 
However, in the case of periodic boundary conditions ($a_1=a_{L+1}$, $b_1=b_{L+1}$), 
which we shall focus on in the following, this duality is only up to degeneracies.

\clearpage
\subsubsection{Bigger (mathematical) picture}
\label{BiggerMathematicalPicture}

\noindent
So far our discussion in this `Supplementary Material'
has been  largely focused on
detailed  algebraic manipulations.
In this subsection we wish to give a brief idea of
the general bigger picture of topological field
theories which underlies these detailed manipulations.
At the same time we will provide a deeper understanding
of the so-called `Levin-Wen model' within this context.

In the main text we have given a physically motivated description of the
Levin-Wen model in Figs.~1 and 2, leading to the Hamiltonian in Eq.~(1) of the main
text. Let us now give a more abstract description of it.
The most general  Levin-Wen Hamiltonian has two kinds of terms: the vertex type (not
discussed so-far as a term of the Hamiltonian)  and plaquette type.  
Let us consider a  surface $\Sigma$, and a 
trivalent graph $\Gamma$  (which we called \lq skeleton\rq in the main text)
embedded in that surface.
(The sole role of the surface $\Sigma$,
which in the leftmost picture of Fig. 1 of the main text
is just a parallel plane sitting in between the two depicted sheets,
is to give a well defined
meaning to the notion of a \lq plaquette\rq; namely,
all complimentary regions of $\Sigma \backslash \Gamma$,
i.e. the complimentary regions of the graph $\Gamma$ within the surface
$\Sigma$, are plaquettes.) 
We always enforce strictly the condition that
three labels meeting at a vertex must satisfy the fusion
rule. (This is another way of saying
that we have set the coupling constant of the `vertex term'
in the most general  Levin-Wen
Hamiltonian to infinity.)
As a result, we obtain a 
Hilbert space called $L(\Gamma, \Sigma)$ consisting of the Hilbert 
space spanned by all admissible labelings of the trivalent graph $\Gamma$: 
a labeling of $\Gamma$ is an assignment of a label in a label set 
$I_{\mathcal{C}}$ to each edge of the graph,
[$I_{\mathcal{C}}=$ $\{{\bf 1}, \tau \} $ in
the previous subsection \ref{SectionFibonacciAnyons}],
and the labeling is admissible if the three labels around each vertex satisfy the  fusion rules.  

Now, there exists another vector space, which brings about the connection
with the actual surfaces that were drawn in Figs.~1 and 2 of the main text.
In particular, when
$\mathcal{C}$ denotes a so-called 
modular category (for a precise definition, 
which we do not need at the moment,
see e.g. Ref. \onlinecite{ZWangPictureTQFT2008})
which basically denotes a theory of `anyons' and their
corresponding  `fusion rules' such as the one 
described in 
the previous subsection \ref{SectionFibonacciAnyons},
then the vector space $L(\Gamma,\Sigma)$ is the same as a Hilbert space
$V_{\mathcal{C}}(S_{\Gamma})$
(for a definition see e.g. Ref. \onlinecite{ZWangPictureTQFT2008})
of an associated  Topological Quantum Field Theory (TQFT)
corresponding to the `modular category'  $\mathcal{C}$:
specifically let $N_{\Gamma}$
be the thickening of  the graph $\Gamma$ to a handle-body (drawing
a cylinder around each edge of the graph), and $S_{\Gamma}$ be the
boundary surface of $N_{\Gamma}$, then $L(\Gamma,\Sigma)\cong V_{\mathcal{C}}(S_{\Gamma})$.  
In the language of TQFT, any `pants-decomposition' 
of the surface  $S_{\Gamma}$ 
is known to lead
to a basis of $V_{\mathcal{C}}(S_{\Gamma})$, which corresponds to the
vector space spanned all possible fusions of the labelings on $\Gamma$. 

This interpretation of the Hilbert space $L(\Gamma,\Sigma)$ gives rise to a transparent derivation 
of the plaquette term, Eq. (\ref{plaq_term}), in the Levin-Wen model.
To derive this expression,
 we use the identification of $L(\Gamma,\Sigma)$ with
$V_{\mathcal{C}}(S_{\Gamma})$.  The $c$th row of the modular $S$-matrix of
the modular category $\mathcal{C}$ can 
be used to construct a projector $\omega_c$ that projects out the particle with a label $c$
through a plaquette. In
other words total flux $c$ through a plaquette $p$ can 
be enforced by inserting $\omega_c$ into a plaquette $p$.  
The projector turns out to be of the form
\begin{equation}
\omega_c
=
{1\over D} \sum_a \ S^a_c \ [a] \,,
\label{OmegaC}
\end{equation}
where $[a]$ denotes a loop labeled by $a$ as
the one drawn  in Eq. (\ref{S_eq}).
In order to see
that this performs the task
let us insert a flux with label $b$ thought the loop $[a]$,
resulting in the  figure drawn
on the l.h.s. of Eq. (\ref{S_eq}),
which we denote in symbols by $[a](b)$.
When we now perform the sum in Eq.~(\ref{OmegaC})
we obtain, upon making use of Eq. (\ref{S_eq}),

\begin{eqnarray}
\omega_c(b)
& :={1\over D} \sum_a \ S^a_c \ \ \   [a](b)
& = 
{1\over D} \sum_a \ S^a_c \ {S^b_a\over S^b_1} \ \  \ [b] 
 = 
\delta_c^b \ {1\over D \ S^c_1} \ \ \  [c]
=  
\delta_c^b \ {1\over d_c} \ \ \  [c]  \\ 
\omega_c\left (\;\; \parbox{3mm}{\scalebox{0.3}[0.3]{\begin{picture}(0,0)%
\includegraphics{Smatrix2.pdf}%
\end{picture}%
\setlength{\unitlength}{3947sp}%
\begingroup\makeatletter\ifx\SetFigFont\undefined%
\gdef\SetFigFont#1#2#3#4#5{%
  \reset@font\fontsize{#1}{#2pt}%
  \fontfamily{#3}\fontseries{#4}\fontshape{#5}%
  \selectfont}%
\fi\endgroup%
\begin{picture}(707,1791)(1168,-994)
\put(1276,464){\makebox(0,0)[lb]{\smash{{\SetFigFont{25}{30.0}{\familydefault}{\mddefault}{\updefault}{\color[rgb]{0,0,1}$b$}%
}}}}
\end{picture}%
}}   \right )
& := 
{1\over D} \sum_a \ S^a_c \; \;\parbox{1.4cm}{\scalebox{0.3}[0.3]{\input{Smatrix1.pdf_t}} }
& =  
\delta_c^b\; \frac{1}{d_c} \;\;\;\parbox{5mm}{\scalebox{0.3}[0.3]{\begin{picture}(0,0)%
\includegraphics{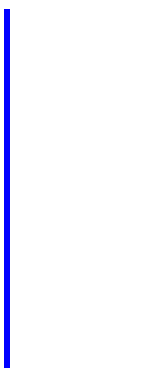}%
\end{picture}%
\setlength{\unitlength}{3947sp}%
\begingroup\makeatletter\ifx\SetFigFont\undefined%
\gdef\SetFigFont#1#2#3#4#5{%
  \reset@font\fontsize{#1}{#2pt}%
  \fontfamily{#3}\fontseries{#4}\fontshape{#5}%
  \selectfont}%
\fi\endgroup%
\begin{picture}(677,1791)(1168,-994)
\put(1276,464){\makebox(0,0)[lb]{\smash{{\SetFigFont{25}{30.0}{\familydefault}{\mddefault}{\updefault}{\color[rgb]{0,0,1}$c$}%
}}}}
\end{picture}%
 }}
\\ \nonumber
\label{OmegaCA}
\end{eqnarray}
where we have used the unitarity (plus reality and symmetry) of
the modular S-matrix, as well
as Eq.s (\ref{QuantumDimensionSMatrix},\ref{TotalQuantumDimensionSMatrix}).

Therefore, the plaquette term $\delta_{\phi(p),1}$ is implemented by inserting the 
projector $\omega_1=\sum_a \frac{d_a}{D^2} \cdot [a]$ into
the plaquette  $p$.
Now the detailed steps leading to
Eq. (\ref{plaq_term}) are easy to understand:
The insertion of $\omega_1$ into the plaquette is
written explicitly in Eq. (\ref{ProjectorImplemented}).
In the subsequent equation,
first an $F$-move is applied to the
two lines connected by the dotted line, and subsequently four more $F$ moves 
counterclockwise around $p$ are implemented as drawn; 
finally removing the resulting bubble, we obtain the
explicit form of the plaquette term
written in Eq. (\ref{plaq_term}).

The mathematical context for the Levin-Wen model is the Drinfeld center $Z(\mathcal{C})$ or 
quantum double of a unitary fusion category $\mathcal{C}$.  The label set $I_{\mathcal{C}}$ for the
Levin-Wen Hamiltonian is the isomorphism classes of simple objects of $\mathcal{C}$.  It is known that a
unitary fusion category is always spherical.  By a theorem of M.~M\"uger \cite{Mueger}, 
the Drinfeld center of any spherical
category is always modular.
It follows that the Drinfeld center of any unitary fusion category is always modular.  
Moreover, if the spherical 
category $\mathcal{C}$ itself is modular, then $Z(\mathcal{C})$ is isomorphic to the 
direct product of the conjugate
$\mathcal{C}^{*}$ and $\mathcal{C}$, where $\mathcal{C}^{*}$ is 
obtained from $\mathcal{C}$ by complex conjugating 
all data.  Our main example is one of those special cases, 
where $\mathcal{C}$ is the Fibonacci theory.  

The decomposition of $Z(\mathcal{C})$
hints directly at the appearance of Dynkin diagram $D_6$ at the critical point
in one-dimensional geometry:  
indeed, the two phases connected by the critical point are the Fibonacci theory and 
the doubled Fibonacci theory with label
sets $\{1,\tau\}$, and $\{(1,1),(1,\tau),(\tau,1),(\tau,\tau)\}$, respectively.  
Based on this, it is natural to expect
that the two sets of fusion rules will fit together in a compatible way at the critical point,
which is nicely illustrated by the structure of the $D_6$ Dynkin diagram in  Fig.  \ref{dynkin}
which underlies the exact solution of this critical point
(Section \ref{SubsectionAnalyticalSolution} below).

\subsection{Topological phases}

We start the detailed discussion of the phase diagram with the 
two distinct gapped non-abelian topological phases: the `single torus' phase where all plaquettes
 are closed at $\theta=0$ and the `two tori' phase with closed off rungs at $\theta=\pi/2$.
A  finite-size scaling analysis of the splitting of the ground state degeneracies and the energy 
gap shows that the phases extend over a wide range of parameter space as illustrated in the phase diagram (Fig. 5b of the main part of the  paper).  In this section we discuss the low-lying excited states in these phases, give their explicit wave functions at the exactly solvable points, and a perturbative expansion for their dispersion away from these points.

\subsubsection{The `single torus' phase at $-\pi/2 < \theta < \pi/4$}

\begin{figure}[t]
\begin{center}
\includegraphics[width=14cm]{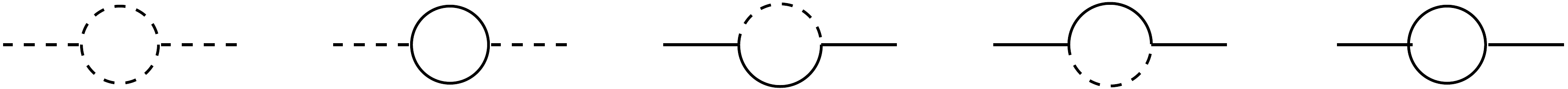}
\caption{Possible basis configurations in the presence  of one plaquette excitation. The two ground states (without a flux through the plaquette) and three excited states (with a $\tau$-flux through the plaquette) are linear combinations of these basis states.
  Solid lines denote $\tau$-anyons while dashed lines symbolize the trivial particle $1$.
}
\label{plaq_conf}
\end{center}
\end{figure}

To describe the lowest excited states we consider the 
trivially solvable point $\theta=0$ where $J_r=0$. 
In the ground state there is no flux through any of the plaquettes, and they all can be closed, thus 
reducing the high-genus ladder
 to a single torus (see Fig. 3 in the main part of the  paper). There are two degenerate ground-states configurations
  with either no flux or a $\tau$-flux through this torus.

  Similarly, we can deduce the degree of degeneracy for the lowest excited state by 
    considering the topology  of this state.
In the lowest excited state, one plaquette flux is present which yields the reduced  topology
(as compared to the high-genus ladder)  and the associated 
 skeleton shown in 
Fig.~4b of the main part of the paper. 
 Closing all but one plaquette this  skeleton  allows for $5$ different $1,\tau$ coverings,
 illustrated schematically in Fig.~\ref{plaq_conf}.
 In order to obtain the  anyon-fluxes through the excited plaquette,  
a	 basis transformation (consisting of a $F$- and a $S$-transformation)  of the reduced basis 
is performed which yields that
 there are three $\tau$-fluxes through each plaquette. Thus, the lowest excited 
 state at $\theta = 0$ is $3L$-fold degenerate.
 Tuning away from $\theta=0$ these $3L$ excitations delocalize and form a three-fold degenerate band.

\subsubsection{The `two tori' phase at $\pi/4 < \theta < \pi$}
At the point $\theta = \pi/2$ (trivially solvable)
the ground state has no $\tau$-anyons on the rungs of the ladder. The rungs can hence be cut
which yields an effective topology of two separate tori. Of the four degenerate ground states three are symmetric and one is antisymmetric under $y$-reflection.
The lowest excited state is a $\tau$-anyon flux through a single rung. The fusion rules then require a flux through both of the two tori, and this state is hence only $L$-fold degenerate. Tuning away from $\theta=\pi/2$, these  states  delocalize into a non-degenerate band.

\subsubsection{Perturbation expansion for the quasiparticle bands}

Over a broad range of parameters the quasi-particle excitations are well described (see Fig. 5b of the 
main part of the paper) by a second order perturbative expansion around $\theta=\pi/2$, with a dispersion given by
\begin{equation}
\Delta E(J_p,J_r,k_x) = J_r- \frac{2J_p}{D^2}\cos(k_x)  -\frac{J_p^2
\varphi}{D^4J_r}[1+2\cos(k_x)]-\frac{J_p^2}{2D^4J_r}2\cos(2k_x).
\label{pert_gap_size}
\end{equation}
Due to duality, this result equally applies for coupling parameters $\theta$ close to $\theta=0$,
 with $J_r$ and $J_p$ interchanged.

\subsection{Gapless theories}

\noindent
In this section, we discuss the critical points ($\theta=\pi/4, 5\pi/4$) and 
the extended critical phase in the ladder model. 
We  first discuss the gapless theories in terms of numerical results and then present
analytical arguments leading to an exact solution for the two critical points ($\theta=\pi/4, 5\pi/4$).

\begin{figure}[b]
\begin{minipage}{0.55\textwidth}
    \includegraphics[width=1\textwidth]{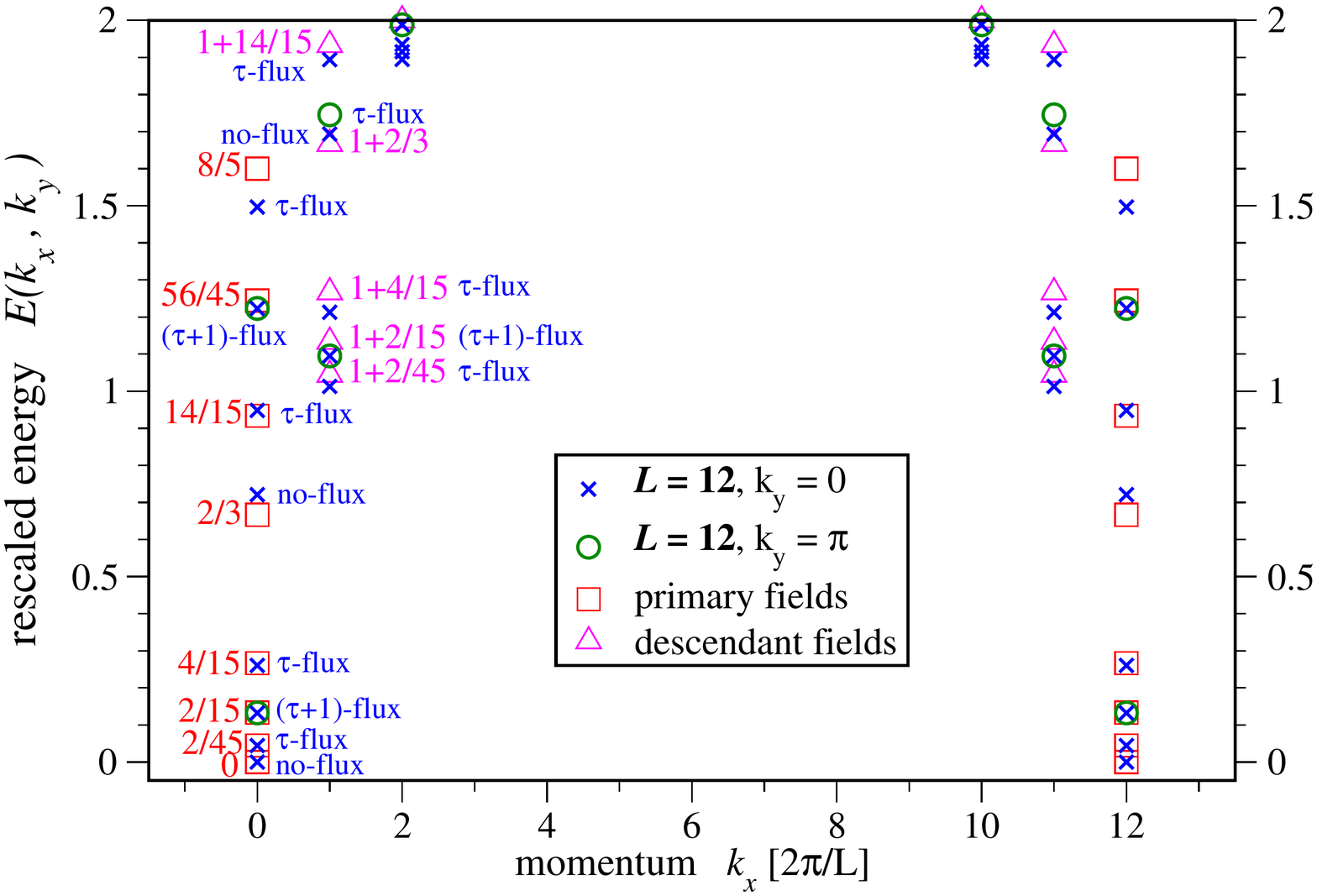}
    \caption{Exact diagonalization:
                    Energy spectrum at the critical point ($\theta=\pi/4$) 
                    for a ladder with $L=12$ holes and 36 anyons.
                    The energies have been rescaled so that the two lowest
                    eigenvalues match the CFT scaling dimensions.
                    The open boxes indicate the primary fields 
of the 7th minimal
          model with central charge $c=14/15$.
                    The topological symmetry sectors are indicated with symbols 
                    $1\equiv y_{1,1}$, $\tau\equiv y_{\tau,\tau}$ and $\tau+1 \equiv y_{1,\tau}$.}
    \label{crit_spec1}
\end{minipage}\hfill
\begin{minipage}{0.4\textwidth}
   \vskip 7mm
    \begin{tabular}{c|c||c|c|c}  
	$(r,s)$ & $h_{(r,s)}+\bar{h}_{(r,s)}$ & $k_x$ & $k_y$ &  $y$\\[2mm] \hline \hline 
	$(1,1)$ & $0$  &$0$ & $0$         & $\varphi^2$   \\[1mm] \hline 
 	$(3,3)$ & $\frac{2}{45}$ &$0$ & $0$  & $\varphi^{-2}$   \\[1mm] \hline 
 	$(5,5)$ & $\frac{2}{15}$  &$0$ & $0$    & $-1$   \\[1mm] \hline 
	$(5,5)$ & $\frac{2}{15}$  & $0$ & $\pi$                          & $-1$   \\[1mm] \hline  
	$(7,7)$ & $\frac{4}{15}$ & $0$ & $0$     & $\varphi^{-2}$   \\[1mm] \hline 
	$(2,1)$ & $\frac{2}{3}$   &$0$ & $0$                         & $\varphi^2$   \\[1mm] \hline  
	$(4,3)$ & $\frac{14}{15}$ &$0$ & $0$                           & $\varphi^{-2}$   \\[1mm] \hline 
	$(6,5)$ & $\frac{56}{45}$ &$0$ & $0$                          & $-1$   \\[1mm] \hline 
	$(6,5)$ & $\frac{56}{45}$  &$0$  & $\pi$                         & $-1$   \\[1mm] \hline 
	$(8,7)$ & $\frac{8}{5}$  &$0$ & $0$                         & $\varphi^{-2}$   
    \end{tabular} 
   \vskip 11mm
    \caption{CFT fields: 
                    Scaling dimensions $h_{(r,s)}+\bar{h}_{(r,s)}$ of the primary fields in the
					 $(D,A)$ modular invariant of the 
                    7th minimal model with central charge $c=14/15$.
                    On the right, we give momentum and topological symmetry assignments
                    of these primary fields for our microscopic model.}
    \label{crit_spec1Top}
\end{minipage}\hfill
\end{figure}

\subsubsection{Critical point at $\theta=\pi/4$ \\(numerical findings from exact diagonalization)}

\label{NumericalFindingsMinimalModel}

At equal positive values of the two coupling constants ($J_p = J_r$,$\theta=\pi/4$), 
the system has a linear 
energy-momentum disperson relation with the finite-size spacing
between energy levels vanishing linearly in $1/L$.
This indicates that the two adjacent, gapped topological phases (Fig. 5) are separated by 
a continuous quantum phase transition and a critical point that is described by a 
2D conformal field theory (CFT).
To characterize this CFT, we rescale and match the
finite-size energy spectra obtained
numerically by exact diagonalization
for systems with up to $L=36$ anyons
to the form of the  spectrum of a CFT,
\begin{equation}
E = E_1L +\frac{2\pi v}{L} \left (-\frac{c}{12}+ h+\bar{h} \right ),
\label{CFT_energy_levels}
\end{equation}
where the velocity $v$ is an 
overall scale factor, and $c$ is the central 
charge of the CFT.
The scaling dimensions $h+\bar{h}$ take the form $h=h^0+n$, $\bar{h}=\bar{h}^0+\bar{n}$,
with $n$ and $\bar{n}$ non-negative integers, and $h^0$ and $\bar{h}^0$ are the holomorphic and antiholomorphic 
conformal weights
of primary fields in a given CFT with central charge $c$. 
The momenta (in units $2\pi/L$) are such that $k_x=h-\bar{h}$ or $k_x=h-\bar{h}+L/2$.
Using this procedure, we
find that for the critical point at $\theta=\pi/4$ the rescaled energy spectrum matches 
the assignments (\ref{CFT_energy_levels})
of part of  the Kac-Table of the $m=9$ unitary Virasoro minimal CFT of
central charge $c=14/15$, 
as shown in Fig.~\ref{crit_spec1}.
In Fig.~\ref{crit_spec1Top}
we list all relevant primary fields of this
CFT
which appear
 and their corresponding scaling 
dimensions.
It turns out that only
the Kac-Table
primary fields $\phi_{r,s}$  with $s=$ odd appear, and those
with $s=5$ have multiplicity two
(the associated  states on the ladder being in the bonding/antibonding 
sectors
of `transverse momenta'
$k_y=0,\pi$),
all others having multiplicity one.
These are precisely those Kac-table primary fields
which occur in the 
so-called (D,A)-modular invariant \cite{CappelliEtAlModular}
of the $m=9$th Virasoro minimal CFT of central charge $c=14/15$.

\begin{figure}
 \begin{center}
   \includegraphics[width=0.75\textwidth]{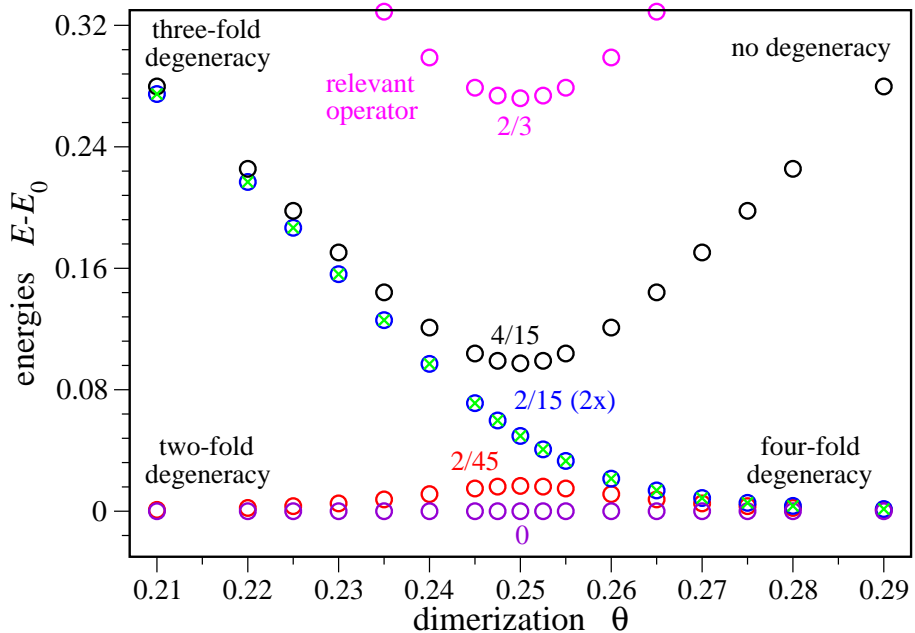}
    \caption{The energies of the lowest lying energy states around the 
                    critical point ($\theta=\pi/4$) as a function of the `dimerization' $\theta$.
                    Results are shown for system size $L=10$.}
   \label{gap_scaling}
\end{center}
\end{figure}

To illustrate how the ground-state degeneracy changes at this critical point from a 
two-fold degeneracy for the `single cylinder' limit ($J_r=0$) to a four-fold degeneracy  
for the `two cylinders'-limit ($J_p=0$), we can follow the evolution of eigenenergies
in the vicinity of this critical point as shown in Fig.~\ref{gap_scaling}.
 Moving away from the critical point ($\theta=\pi/4$) corresponds to a dimerization of the model: in 
  an alternative basis choice, discussed in
   detail in section \ref{SubsectionAnalyticalSolution}, it becomes 
   apparent that the rung and plaquette terms alternatingly act on
   even and odd `sites'.
For $\theta \searrow \pi/4$, the four-fold ground-state degeneracy is lifted with one 
of the four ground states approaching the field with rescaled energy $2/45$ ($k_y=0$), 
and two degenerate ground states moving to a rescaled energy $2/15$ 
($k_y=0$ and $k_y=\pi$).
The single first excited state in this gapped phase softens towards the rescaled 
energy $4/15$ at the critical point.
As we move into the adjacent gapped phase for $\theta < \pi/4$ only the field with
rescaled eigenenergy $2/45$ moves back towards the ground-state, while the two other 
fields move upwards in energy and form a three-fold degenerate excited state.

\begin{figure}
\begin{minipage}{0.55\textwidth}
    \includegraphics[width=1\textwidth]{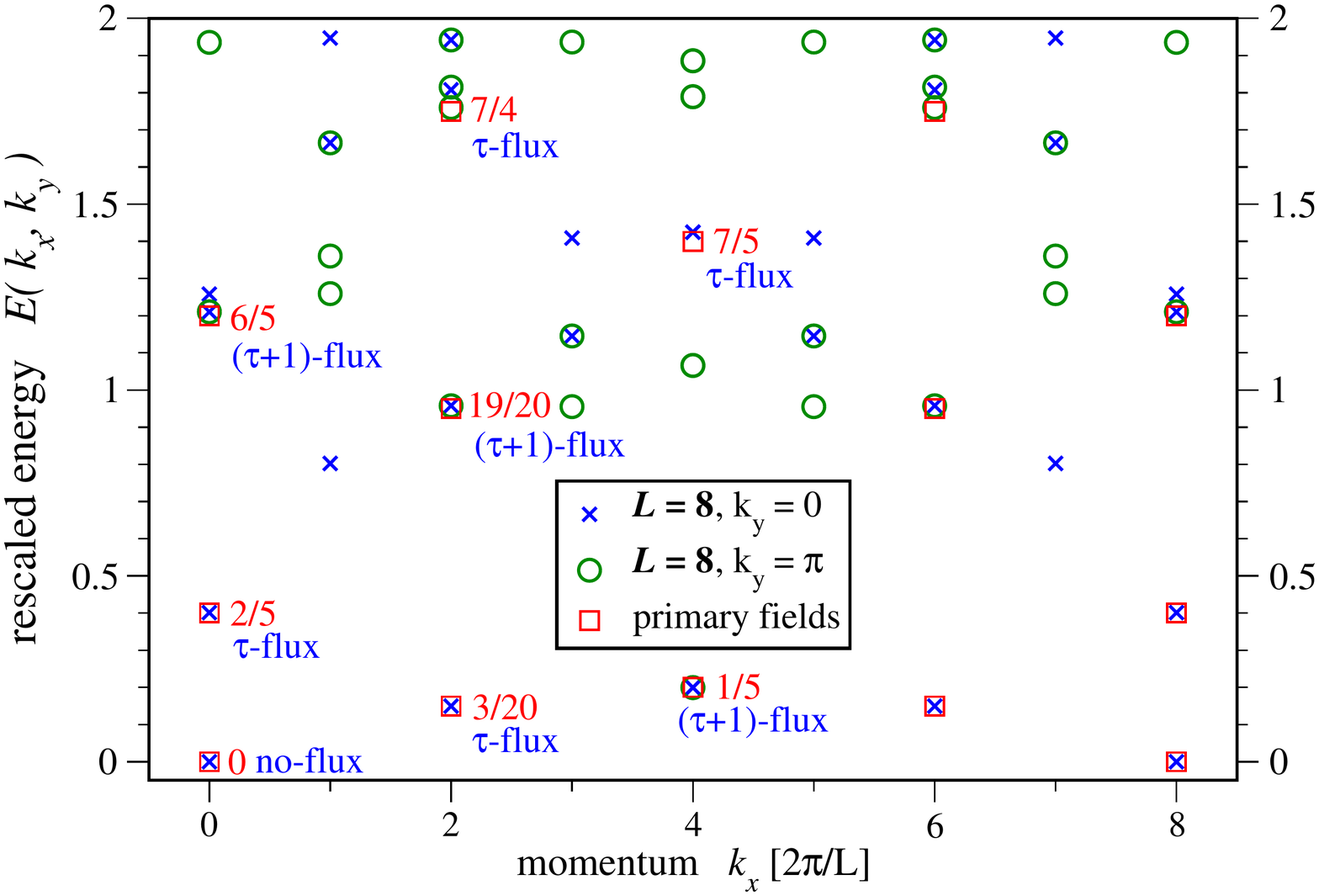}
    \caption{Exact diagonalization: 
                    Energy spectrum at the critical point ($\theta=5\pi/4$) 
                    for a ladder with $L=8$ holes and 24 anyons.
                    The energies have been rescaled so that the two lowest
                    eigenvalues match the CFT scaling dimensions.
                    The open boxes indicate the primary fields of the $Z_8$-parafermion
                    model with central charge $c=7/5$.
                    The topological symmetry sectors are indicated with symbols 
                    $1\equiv y_{1,1}$, $\tau\equiv y_{\tau,\tau}$ and $\tau+1 \equiv y_{1,\tau}$.}
    \label{crit_spec2}
\end{minipage}\hfill
\begin{minipage}{0.4\textwidth}
   \vskip 7mm
    \begin{tabular}{c|c||c|c|c}  
   	$(j,m)$ & $h_{(j,m)}+\bar{h}_{(j,m)}$ &$k_x$& $k_y$ & $y$\\ \hline \hline 
 	$(0,0)$ & $0$       &   $0$     & $0$                        & $\varphi^2$   \\[1mm] \hline 
	$(1,1)$ & $\frac{3}{20}$ &    $\frac{\pi}{4}$     & $0$       & $\varphi^{-2}$   \\[1mm] \hline  
 	$(2,2)$ & $\frac{1}{5}$  &    	$\frac{\pi}{2}$     & $0$                  & $-1$   \\[1mm] \hline 
 	$(2,2)$ & $\frac{1}{5}$   &   	$\frac{\pi}{2}$     & $\pi$         & $-1$   \\[1mm] \hline 
	$(1,0)$ & $\frac{2}{5}$     &     	$0$     & $\pi$     &     $\varphi^{-2}$   \\[1mm] \hline  
	$(2,1)$ & $\frac{19}{20}$    &   	$\frac{\pi}{4}$     & $0$       & $-1$   \\[1mm] \hline
	$(2,1)$ & $\frac{19}{20}$     & $\frac{\pi}{4}$     & $\pi$            & $-1$   \\[1mm] \hline 
 	$(2,0)$ & $\frac{6}{5}$      &          $0$     & $0$                   & $-1$   \\[1mm] \hline 
	$(2,0)$ & $\frac{6}{5}$          &     $0$     & $\pi$                       & $-1$   \\[1mm] \hline 
   	$(3,2)$ & $\frac{7}{5}$  &   $\frac{\pi}{2}$     & $0$                 & $\varphi^{-2}$   \\[1mm] \hline 
  	$(4,3)$ & $\frac{7}{4}$             &  	$\frac{\pi}{4}$     & $0$                 & $\varphi^{-2}$  
    \end{tabular} 
   \vskip 9mm
    \caption{CFT fields: 
                    Scaling dimensions $h_{(j,m)}+\bar{h}_{(j,m)}$ of the primary fields in the 
                    $Z_8$-parafermion CFT with central charge $c=7/5$.
                    On the right, we give momentum and topological symmetry assignments
                    of these primary fields for our microscopic model.}
    \label{crit_spec2Top}
\end{minipage}\hfill
\end{figure}

\subsubsection{Extended critical phase for $\theta \in (\pi,3\pi,2)$\\
(numerical findings from exact diagonalization)}
\label{NumericalFindingsParafermions}

For negative coupling parameters $J_p, J_r < 0$, we find an extended critical phase around
the point of equal coupling strength which in our circle phase diagram is opposite to the critical 
point discussed above.
For the whole extent of this critical phase we can match
the finite-size energy spectra
to the $Z_8$ parafermion CFT with central charge $c=7/5$.
This theory is part of the sequence of $Z_k$-parafermion  CFTs with conformal weights 
$\Delta^j_m= {j(j+1)\over k+2} - {m^2\over k}$, where $j=0,\frac{1}{2},1,...,k/2$, $|m|\le j$ 
(and $j-m=$ integer),
in the notation of \cite{ZamolodchikovFateevCurrentAlgebraZk_1}.
The details of the assignments for $k=8$ can be found in Fig.~\ref{crit_spec2} and Table II.

In order to verify that the critical phase around the exactly soluble point $\theta=5\pi/4$ 
extends to the vicinity of the decoupling points $\theta=\pi$ and $\theta=3\pi/2$, 
we consider an effective model where we fix all rung occupations to $\tau$-anyons. 
This assumption is exactly true at the decoupling point $\theta=3\pi/2$. 
Implementing this constraint significantly reduces the size of the Hilbert space and allows
us to numerically study this effective model for larger system sizes with up to 48 anyons.

\noindent
The effective Hamiltonian in the reduced Hilbert space is  given by
\begin{equation}
H^{\rm eff} = -LJ_r - J_p \sum_{{\rm plaq} \ p} \delta_{\phi(p),1} \, .
\label{Ham_eff}
\end{equation}
The first term is a constant, and can thus be omitted which then turns the actual value of $J_p$   irrelevant. A
 positive $J_p$ corresponds to the limit $\theta\searrow 3\pi/2$, while a negative
$J_p$ allows to study the limit $\theta \nearrow 3\pi/2$.

\begin{figure}[h]
  \begin{center}
 \includegraphics[width=0.7\textwidth]{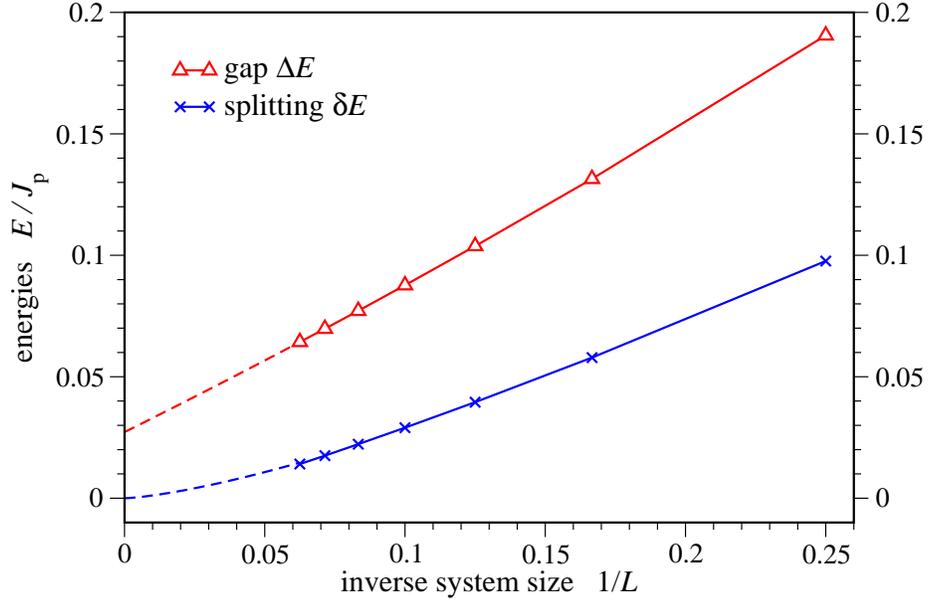}
 \caption{Energy gap $\Delta E(1/L)/J_p$ between the first excited state and the  ground state,
  as well as the splitting of the ground state degeneracy, $\delta E(1/L)/J_p$, 
 for the effective model Eq. \ref{Ham_eff}. The two ground states
 become precisely degenerate only in the 
 thermodynamic limit. The  results indicate that the energy gap extrapolates to a finite value. 
  Since the effective model is valid in the limit $\theta\rightarrow3\pi/2+$,
   the gapped topological phase extends all the way up to this point.}  
 \label{gap_scaling2}
 \end{center}
\end{figure}

For positive $J_p$, we find that the splitting of the ground state degeneracies goes to zero for 
$1/L\to 0$, and the energy gap approaches a finite value as shown in Fig.~\ref{gap_scaling2}.
This further supports the stability of the gapped topological phases up to, but excluding, 
the points $\theta=\pi$ and $\theta=3\pi/2$ in our phase diagram. 

For negative $J_p$, the rescaled energy spectrum of this effective model is critical and again
matches (with much higher accuracy than at $\theta=\pi$)
the $Z_8$ parafermion conformal field theory with central charge $c=7/5$ 
as shown 
in Fig.~\ref{crit_spec2}.
We can hence conclude that  the whole quadrant $\theta \in (\pi,3\pi/2)$ is occupied by an
extended critical phase described by the same conformal field theory as the exactly solvable
point $\theta = 5\pi/4$.

Approaching the endpoints of this extended critical phase at $\theta=\pi$ and $\theta=3\pi/2$, 
the low-energy spectrum collapses into a flat band resulting in an extensive ground state 
degeneracy below an energy gap of size $1$ at the points $\theta=\pi$ and $\theta=3\pi/2$.
Moving beyond these `decoupling points' where one of the terms in the Hamiltonian vanishes,
this extensive ground-state degeneracy is split again and a gap opens for $\theta<\pi$ and
$\theta>3\pi/2$, respectively, as the system enters the two gapped, topological phases discussed
above.

\begin{figure}[h]
  \begin{center} 
   \includegraphics[width=0.7\textwidth]{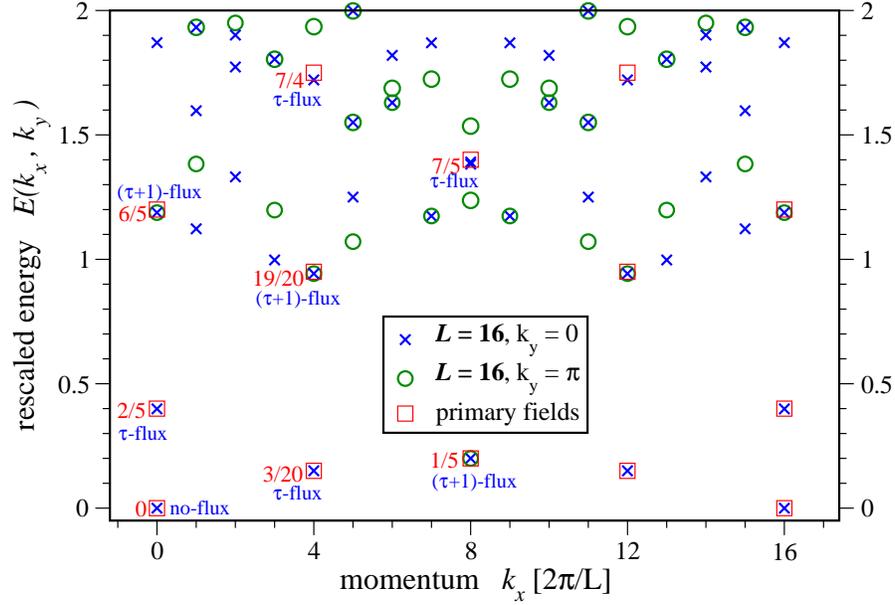}
   \caption{Rescaled energy spectrum of the effective
 	          model with $J_p$ negative ($L=16$), and $Z_8$ parafermion CFT assignments.
                    The topological symmetry sectors are  indicated with symbols 
                    $1$ ($y_{1,1}=\varphi^2$), $\tau$ ($y_{\tau,\tau}=\varphi^{-2}$) and $\tau+1$         
                    ($y_{1,\tau}=-1$).}
\label{crit_spec3}
\end{center} 
\end{figure}

\clearpage
\subsubsection{Topological stability of the  critical phases}

\begin{figure}[b]
\begin{center}
 \scalebox{0.4}[0.4]{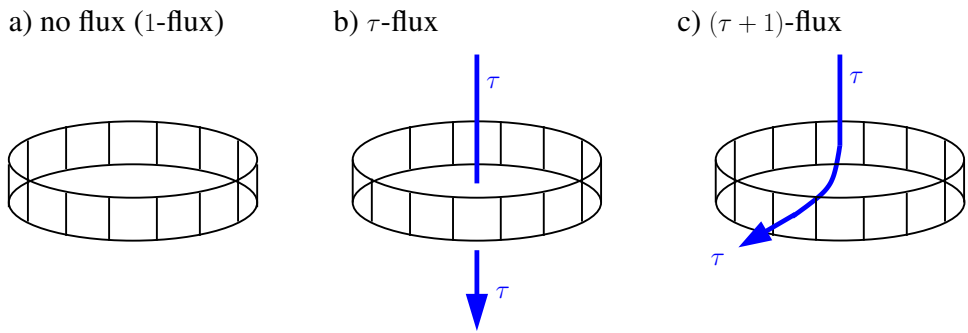}
 \caption{Topological symmetry sectors: 
 	       a) No $\tau$-flux enters or leaves the ladder.
		   	       b)  A $\tau$-flux enters from one side and leaves at the other side. 
	       c) A $\tau$-flux enters from one side and leaves through a plaquette. }
 \label{top_sym}\end{center}
\end{figure}

Both critical theories have  a
large number
 of  rescaled energies 
(\ref{CFT_energy_levels}) that are smaller than  two.
These eigenenergies are associated with 
operators whose correlation functions
decay with scaling exponent $h+\bar{h}<2$.
Such operators are relevant in the
 renormalization group sense.  
This means that any operators
${\cal O}$ with scaling dimensions (=rescaled energies)
$h+\bar{h}<2$ 
which is invariant under all symmetries of the Hamiltonian 
may appear as an additional term in the latter and can thus
drive the system out of the critical phase into a gapped phase or a 
different critical phase.
For a critical phase to be stable there must hence exist  a symmetry in the model such 
that the identity field (associated with the ground state) belongs to a different symmetry 
sector than all other fields $\phi$ with $h+\bar{h}<2$ and $k_x=0$
(fields at $k_x\ne 0$ do not obey the translational symmetry of the Hamiltonian).
Indeed, our model has an additional topological symmetry \cite{GoldenChain} that
can stabilize the critical phases: 
There can be either no flux (denoted as $1$-flux) or a $\tau$-flux entering the periodic 
ladder from one side, and a $1$- or a $\tau$-flux  leaving the ladder as illustrated in 
Fig.~\ref{top_sym}. 
There are hence three possibilities for possible flux assignments: 
(i) no flux is entering  from above, and no flux is leaving [Fig.~\ref{top_sym}a],  
(ii) a $\tau$-flux is entering and leaving [Fig.~\ref{top_sym}b], or, 
(iii) a $\tau$-flux is entering from one side, and leaves through one or several plaquettes
as shown in Fig.~\ref{top_sym}c).
For each operator, one of the three scenarios applies and we can explicitly determine the
topological sectors by considering the following hermitian symmetry operator 
(which commutes with the Hamiltonian)
 \begin{align}
Y|a,b,c \rangle & = \sum_{a',b'}\; \prod_{i=1}^{L}(F_{c_i a_i \tau}^{a_{i+1}'})_{a_i'}^{a_{i+1} } 
(F_{c_i b_i \tau}^{b'_{i+1}})_{b_i'}^{b_{i+1}} |a',b',c\rangle \,,
\end{align}
where $|a,b,c\rangle = |a_1,b_1,c_1,a_2,b_2,c_2,....,a_L,b_L,c_L\rangle$ are 
labels according to Fig.~\ref{ladder_surface}. 
This operator inserts additional $\tau$-loops parallel to the two `spines' of the ladder. 
As in the case of the plaquette term Eq.~(\ref{plaq_term}), this is done by connecting them 
to the ladder with $1$-particles.
The flux through each of  these two additional $\tau$-loops
can be either $1$ or $\tau$, where a $1$-flux yields a factor of $S_1^{\tau}/S_1^1 = \varphi$, 
and a $\tau$-flux gives $S_{\tau}^{\tau}/S_{\tau}^1 = -\varphi^{-1}$ (note that a $S$-transformation has
to be performed in order to obtain the flux through the additional $\tau$-loops).
Hence there are three possible eigenvalues of $Y$:
$y_{1,1}=\varphi^2$ (scenario i), $y_{\tau,\tau}=\varphi^{-2}$ (scenario ii) 
or $y_{1,\tau} = -\varphi^{-1}\varphi =-1$ (scenario iii).

We numerically evaluate the topological symmetry sectors  in the two critical phases
(see Tables I and II, and Figs.~\ref{crit_spec1}, \ref{crit_spec2} and \ref{crit_spec3}). 
At the critical point separating the topological phases ($\theta=\pi/4$), we find
that the relevant operators can be classified according to $s=1 \leftrightarrow y_{1,1}$, 
$s=3,7\leftrightarrow y_{\tau,\tau}$, $s=5\leftrightarrow y_{1,\tau}$.
In particular, only one operator, $\phi_{(2,1)}$, is in the same topological symmetry sector 
as the ground state, i.e. the identity field $\phi_{(1,1)}$. 
It is this field that drives the system out of the critical phase when varying the coupling constant 
$\theta$. With the scaling dimension of this operator 
being $x=2/3$ the
gap opens as 
$\Delta E(\theta) \propto|\theta-\pi/4|^{\nu}$ on either side of the critical point,
where
$\nu = 1/(2-2/3)=3/4$.
In the second critical phase, $\theta\in (\pi,3\pi/2)$, the topological symmetry assignments of 
the relevant operators are given by $r=0 \leftrightarrow y_{1,1}$, $r=2,6\leftrightarrow y_{\tau,\tau}$, $r=4\leftrightarrow y_{1,\tau}$.
In particular, there is no relevant field in the same topological symmetry sector as the ground state,
which implies that there is no symmetry-allowed relevant operator in this gapless theory and the
critical point must be part of an extended gapless phase. This observation 
demonstrates that our observation
(from exact diagonalization studies)
that the extended critical phase in the quadrant $\theta\in (\pi,3\pi/2)$ is described by
the same $Z_8$ parafermion CFT with central charge $c=7/5$
is correct.


\clearpage
\subsection{Analytical solution}
\label{SubsectionAnalyticalSolution}

\noindent
Our ladder model defined by the Hamiltonian in Eq.~(2) in the main part 
of the paper can be {\em solved exactly} at the two critical points 
$\theta=\pi/4$ and $\theta=5\pi/4$ 
(see the phase diagram in Fig.~5 of the main text).
The key observation leading to this exact solution is that the topological
structure of our model implies that its Hilbert space is in fact built on the 
so-called $D_6$-Dynkin diagram, which is drawn below in 
Fig.~\ref{dynkin}.

\begin{figure}[h]
  \begin{center}
	\includegraphics[width=6cm]{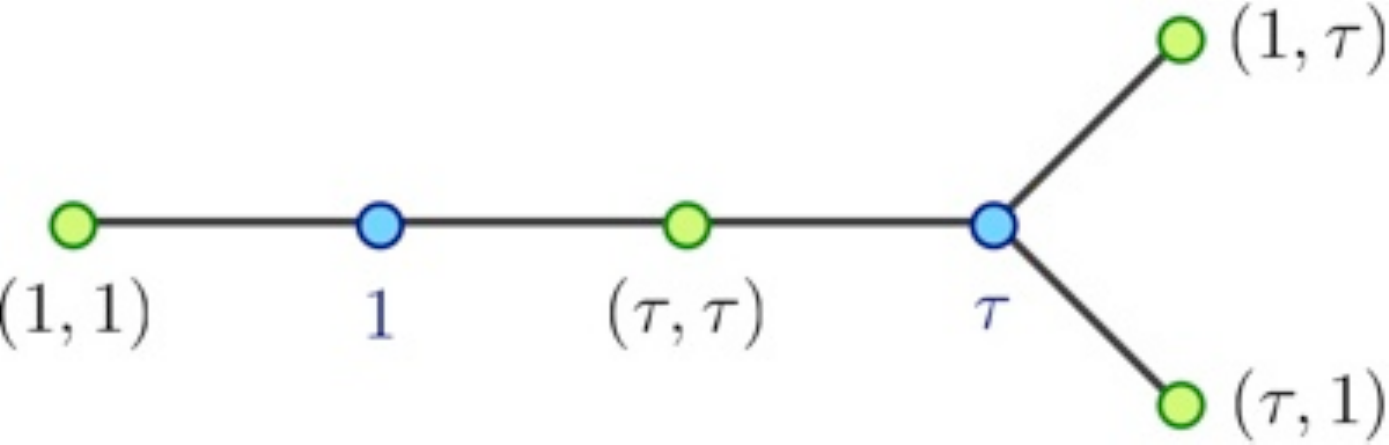}
  \end{center}
  \caption{Dynkin diagram $D_6$.}
  \label{dynkin}
\end{figure}

\begin{figure}[h]
\vskip4mm
\begin{center}
\scalebox{0.42}[0.42]{\input{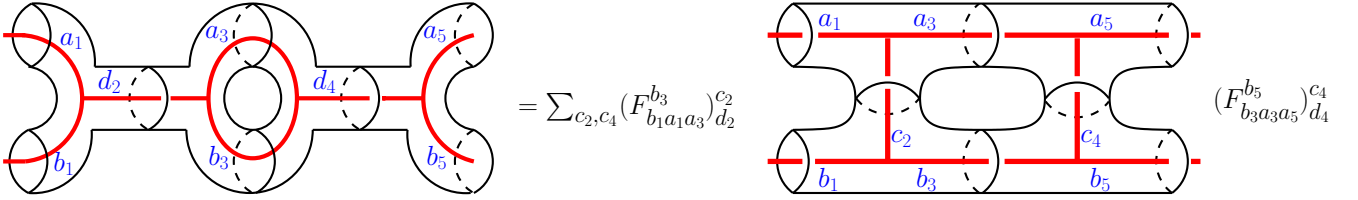}}
\end{center}
\caption{Two possible basis choices corresponding to different decompositions of the
	       high-genus surface. 
                The basis drawn on the left is used in formulating the exact solution: 
                the rung and the plaquette terms alternatingly act on even or odd \lq sites\lq\  $i$.}
\label{second_basis}
\end{figure}

The Dynkin diagram indeed appears very naturally:
let us make a
change of basis for our Hamiltonian
as illustrated in Fig.~\ref{second_basis}. 
This new choice of basis (drawn on the left), 
which arises from a different decomposition of the high-genus surface, 
is related to the original one (drawn on the right) by a simple $F$-transformation.
In particular, consider the new basis in the left part of Fig.~\ref{second_basis}:
with the even-numbered
\lq sites\rq\  (which correspond to the original rungs) 
we associate a  label $d_i=1$ or $d_i=\tau$ 
(the flux through that cross-section of the surface).
With the odd-numbered \lq sites\rq\  (which correspond to the original plaquettes) 
we associate
a variable consisting of a pair of labels, $(a_i,b_i)$ which 
can assume four values, i.e., $(a_i,b_i)=(1,1)$, $(a_i,b_i)=(\tau,1)$, $(a_i,b_i)=(1,\tau)$
and $(a_i,b_i)=(\tau,\tau)$,
and denotes the pair of fluxes through the
two cross-sections of the surface at the position of the plaquette.
The allowed fusion channels at the vertices where variables $(a_i,b_i)$ and $d_{i\pm 1}$ 
meet then correspond precisely to the condition
that they be adjacent nodes on the Dynkin diagram of the  $D_6$ Lie algebra,
as illustrated in Fig.~\ref{dynkin} above.
For example, a local label $(a_i,b_i)=(\tau,\tau)$
at an odd-numbered \lq site\lq\  $i$  allows for labels $d_{i-1}=1$ and $d_{i-1}=\tau$
at the neighboring even-numbered sites,
which is reflected in the fact that label $(\tau,\tau)$ is connected by a line to both labels
$1$ and $\tau$ in the Dynkin diagram.

In summary, the elements of this new basis of the Hilbert space
on which the Hamiltonian acts are of the form
\begin{equation}
\label{RSOSbasisD6}
\ket{{\vec \alpha}} := \ket{ \ldots ,\alpha_{i-1}, \alpha_i, \alpha_{i+1}, \ldots} \,,
\end{equation}
where $\alpha_j$ [$=d_j$ if $j$ is even, and $=(a_j, b_j)$ if
$j$ is odd]  denotes a  point on the $D_6$-Dynkin diagram
representing the flux through the high-genus surface at the \lq site\rq \ $j$ of the chain.
The sequence of $\alpha_j$ must satisfy the condition that
$\alpha_{j+1}$ is a nearest neighbor site of $\alpha_j$
on the $D_6$-Dynkin diagram.

In this new basis,
the rung and plaquette terms
$ H^R_{i}$ and $H^P_{i}$
of our ladder
Hamiltonian
 \begin{align}
\label{Ham2}
 H &= -
J_p
\sum_{i\;{\rm odd}}
 H^P_{i}
-
J_r
\sum_{i\;{\rm even}}
  H^R_{i}\; ,
 \end{align}
take on the following 
form \cite{FootnotePlaquetteTerm}
\begin{align}
\nonumber
H^P_i|a_i,b_i\rangle & = \sum_{s=1,\tau}\;\frac{d_s}{D^2} \sum_{a_i',b_i'} (F^{a_i'}_{ d_{i+1}b_i s })_{a_i
}^{b_i'}
(F^{b_i'}_{d_{i-1} a_i s})_{b_i}^{a_i'} |a_i',b_i'\rangle \,,
\\
\label{Ham2a}
H^R_i|d_i\rangle &= \sum_{d'_i} (F_{b_{i-1}a_{i-1}a_{i+1}}^{b_{i+1}})_{d_i}^1
(F_{b_{i-1}a_{i-1}a_{i+1}}^{b_{i+1}})_{d'_i}^1  |d'_i\rangle \,.
 \end{align}
In fact, these terms
can be seen to form
a representation
of the Temperley-Lieb  algebra \cite{TemperleyLieb}
which arises from the $D_6$-Dynkin diagram,
and has ``d-isotopy"  parameter $D=$
$\sqrt{1+\varphi^2}=$ $2 \cos(\pi/10)$,
the total quantum dimension of our Fibonacci theory.
Specifically, consider the operators ${\bf e}_i$ constructed from the
components $v_\alpha=\sin(\alpha \pi/10)$ ($\alpha=1, ..., 6$, $v_1=v_{(1,1)}$, 
 $v_2=v_{(1)}$, $v_3=v_{(\tau,\tau)}$, $v_4=v_{(\tau)}$, $v_5=v_6=v_{(1,\tau)}=v_{(\tau,1)}$)
of the (\lq Perron Frobenius\rq) eigenvector corresponding to $D$,
the  largest positive 
eigenvalue of the adjacency matrix of the $D_6$-Dynkin 
diagram \cite{FootnoteAdjacencyMatrix},
\begin{align}
\nonumber
{\bf e}_i \ \ket{ \ldots ,\alpha_{i-1}, \alpha_{i}, \alpha_{i+1}, \ldots}
& :=
\sum_{{\alpha'}_i}
\ \ ((e_i)_{\alpha_{i-1}}^{\alpha_{i+1}})^{{\alpha'}_i}_{\alpha_i}
\ \ket{ \ldots ,\alpha_{i-1}, {\alpha'}_{i}, \alpha_{i+1}, \ldots},
\\ 
\label{ei}
{\rm
where
} 
\qquad
\quad
((e_i)_{\alpha_{i-1}}^{\alpha_{i+1}})^{{\alpha'}_i}_{\alpha_i}
 &= \delta_{\alpha_{i-1},\alpha_{i+1}  }
\sqrt{
\frac{v_{\alpha_i} v_{{\alpha'}_i}}
{v_{\alpha_{i-1}} v_{\alpha_{i+1}}}
} \,.
\end{align}
These operators form a  known representation~\cite{PasquierNPB1987} 
of the Temperley-Lieb algebra
with ``d-isotopy"- parameter $D$,
i.e.
\begin{equation}
{\bf e}_i^2=D \  {\bf e}_i \,,
\quad
{\bf e}_i  {\bf e}_{i\pm1} {\bf  e}_i = {\bf  e}_i \,,
\qquad 
[{\bf e}_i, {\bf e}_j]=0 \quad {\rm for} \ |i-j|\geq 2 \,.
\label{TemperleyLiebRelations}
\end{equation}
Now one can check that the rung and plaquette terms,
Eq.~(\ref{Ham2a}), 
of the Hamiltonian
in the new basis, Eq.~(\ref{Ham2}),
are proportional to these operators, i.e.
\begin{equation}
\label{HRandHP}
H^P_{i}  =
{1 \over D} \  \ {\bf  e}_i  \ \  {\rm for}\  i \ {\rm   odd},
\qquad 
H^R_{i}  = 
 {1\over D} \ \   {\bf e}_i  \ \  {\rm for}\  i \ {\rm   even}.
\end{equation}

The Hamiltonian
Eq.~(\ref{Ham2a})  is in fact that corresponding to the
(integrable) restricted-solid-on-solid (RSOS) statistical mechanics 
lattice model based on the $D_6$-Dynkin diagram
\cite{PasquierNPB1987}.
Specifically, the two-row transfer matrix ${\bf T}:={\bf T}_2 {\bf T}_1$ 
of this lattice model
\begin{center}
\includegraphics[width=0.7\textwidth]{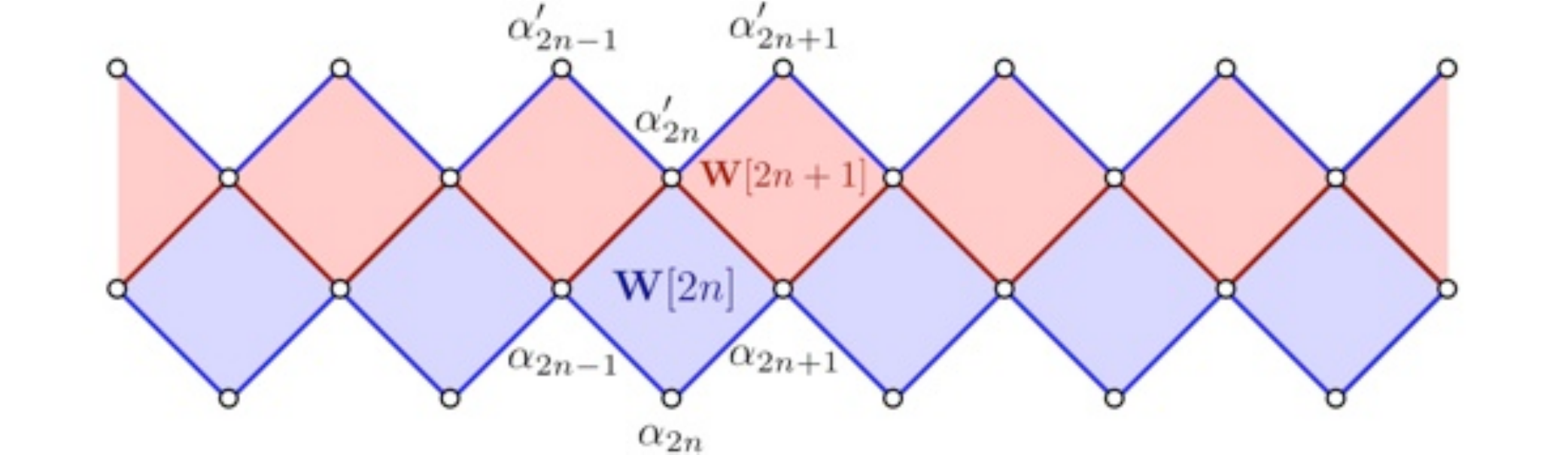}
\end{center}
is written in terms 
of Boltzmann weights ${\bf W}[i]$ assigned to a plaquette $i$ of the square lattice
\begin{equation}
{\bf T}_1 := \prod_{n} {\bf W}[2n]\,, \quad {\rm and} \quad
\ \ \ 
{\bf T}_2 := \prod_{n} {\bf W}[2n+1]
\end{equation}
with
\begin{equation}
\label{DefWi}
{\bf W}[i]^{\vec {\alpha'}}_{\vec \alpha}
=
\left\{
{
\sin[ {\pi\over 10} - u]\over \sin{\pi\over 10}
}
\ 
{\bf 1}^{\vec {\alpha'}}_{\vec \alpha}
+
{ 
\sin{u} 
\over 
\sin[{\pi\over 10}] 
}
{\bf e}[i]^{\vec {\alpha'}}_{\vec \alpha}
\right\} \,.
\end{equation}
The parameter $u>0$ is a measure of the lattice anisotropy,
${\bf 1}$~is the identity operator, and
\begin{equation}
{\bf e}[i]^{\vec {\alpha'}}_{\vec \alpha}
:= \ \ 
\label{ProjectorsRow}
 \left[\prod_{m\not = i} \delta_{{\alpha'}_m,\alpha_m}\right]
\ \ \left({({\bf e}_i)}^{\alpha_{i+1}}_{\alpha_{i-1}}\right)_{\alpha_i}^{{\alpha'}_i} \,.
\end{equation}

The Hamiltonian of the so-defined lattice model
is obtained from its transfer matrix by taking, 
as usual \cite{BaxterBook},
the extremely anisotropic limit, $0 < u \ll 1$,
\begin{equation}
\nonumber
{\bf T} =
\exp\{ - a ({\bf H}+c_1) + O(a^2) \},
\ \ \ a = {u \over D \ \sin[\pi/10]} \ll 1\; ,
\end{equation}
yielding
\begin{equation}
\label{Hamiltonian-ei}
H
= - \sum_i \ \  \frac{1}{D} \ {\bf e_i} \; .
\end{equation}
Since, due
 to Eq.~(\ref{HRandHP}),
the operators  \lq $\frac{1}{D} \ {\bf e_i}$\rq  \ 
are nothing but the rung and plaquette
operators,
we have
thus demonstrated that
the Hamiltonian of the RSOS statistical mechanics model
based on the  Dynkin diagram $D_6$
coincides with the Hamiltonian,
Eq.~(\ref{Ham2}), of our ladder model.

The RSOS model based on $D_6$ is 
known \cite{PasquierIntegrable,PasquierDAPartitionFunction}
to provide
an  (integrable)
lattice realization
of the $(D,A)$ modular invariant
\cite{CappelliEtAlModular}
of the 7th unitary
minimal CFT of central charge $c=14/15$.
In particular, the Hamiltonian of Eq. (2)
of the main text
at angle $\theta=\pi/4$ will yield
the spectrum of that CFT.
This exact analytical result is borne
out precisely by  our numerical (exact diagonalization)
studies reported in subsection
(\ref{NumericalFindingsMinimalModel}).
This CFT with central charge $c=14/15$ describes the quantum critical point of a $1+1$~D 
quantum system, our ladder model. While we cannot make an exact statement for the related
$2+1$~D quantum model, 
we note that Fendley has recently discussed a $2+1$~D quantum critical point 
from a $2+0$~D perspective \cite{Fendley} 
by considering a one-parameter family of wavefunctions connecting
the ground-state wavefunctions of the two extreme limits of the $2+1$~D model \eqref{eq:Hamiltonian}.
For a certain value of the parameter he finds a conformal quantum critical point whose ground-state correlators are written in terms of this same $c=14/15$ CFT.

Another version of this lattice model yielding in the anisotropic limit
the negative, $- H$, of the
Hamiltonian in Eq. (\ref{Hamiltonian-ei})
is also integrable and 
provides \cite{Kuniba}
a lattice realization
of the $Z_8$ parafermionic CFT of central charge $c=7/5$.
In particular, the Hamiltonian of Eq. (2)
of the main text
at angle $\theta=5\pi/4$ will yield
the spectrum of that CFT.
Again, this exact analytical result is 
borne
out precisely  by  our numerical (exact diagonalization)
studies reported in subsection
(\ref{NumericalFindingsParafermions}).

\section{The honeycomb lattice model}
 In this section, we  discuss details of the ``honeycomb lattice model" whose Hamiltonian is given by
 Eq.~(1) in the paper.
 We first define the plaquette term of the model and then discuss  two limiting phases of the model.
\subsection{The Hamiltonian}
In analogy to the plaquette term in the ladder model, Eq.~(\ref{plaq_term}),
  the plaquette term of the honeycomb lattice model (Eq.~(1) in the main text) is defined  by
  \begin{equation}
\delta_{\phi(p),1} \left | 
\parbox{2.cm}{ \scalebox{0.32}[0.32]{
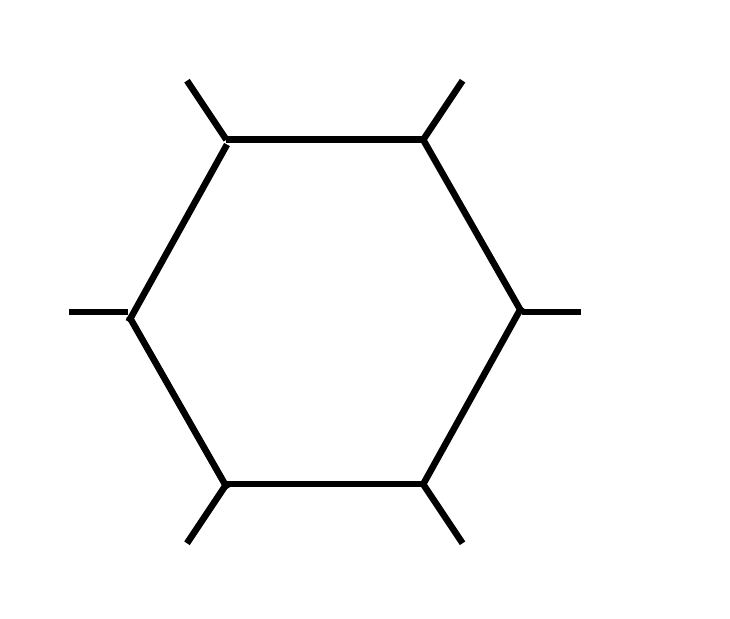 }}
 \right \rangle 
=\sum_{s=1,\tau}\frac{d_s}{D^2} \sum_{\stackrel{\alpha',\beta',\gamma'}{\delta',\epsilon',\zeta'}}
 (F_{a \alpha' s}^{\zeta})_{\alpha}^{\zeta'}
 (F_{f \zeta' s  }^{\epsilon})_{\zeta}^{\epsilon'} 
(F_{e \epsilon' s}^{\delta})_{\epsilon}^{\delta'} 
 (F_{d \delta' s }^{\gamma})_{\delta}^{\gamma'}
 (F_{c \gamma' s}^{\beta})_{\gamma}^{\beta'} 
(F_{b  \beta' s}^{\alpha})_{\beta}^{\alpha'}
\left |  
\parbox{2.cm}{ \scalebox{0.32}[0.32]{
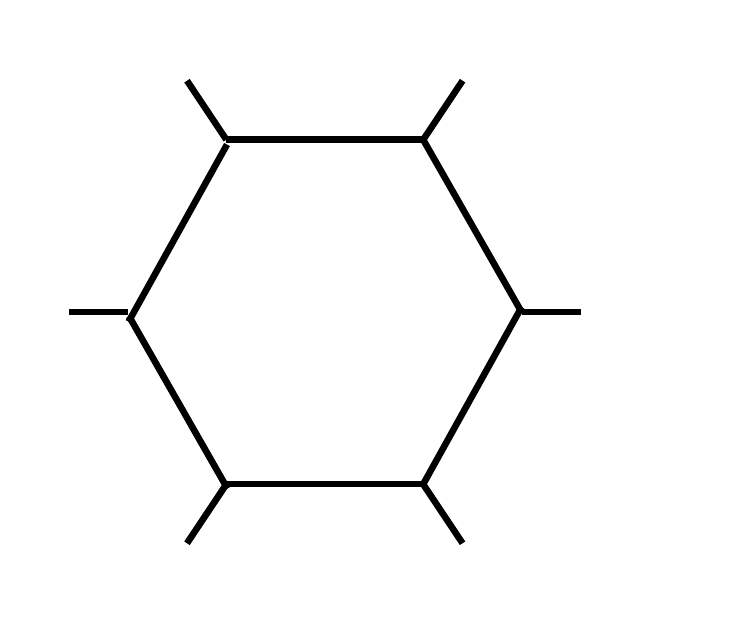 }}
 \right \rangle 
\end{equation}
where the additional two edges of a plaquette are reflected in 
 two additional $F$-transformations.
  Again, we can parametrize the coupling constants on a circle as $J_p=\cos(\theta)$ and $J_e = \sin(\theta)$.
 
\subsection{Excitations}

\begin{figure}[ht]
\includegraphics[width=0.35\textwidth]{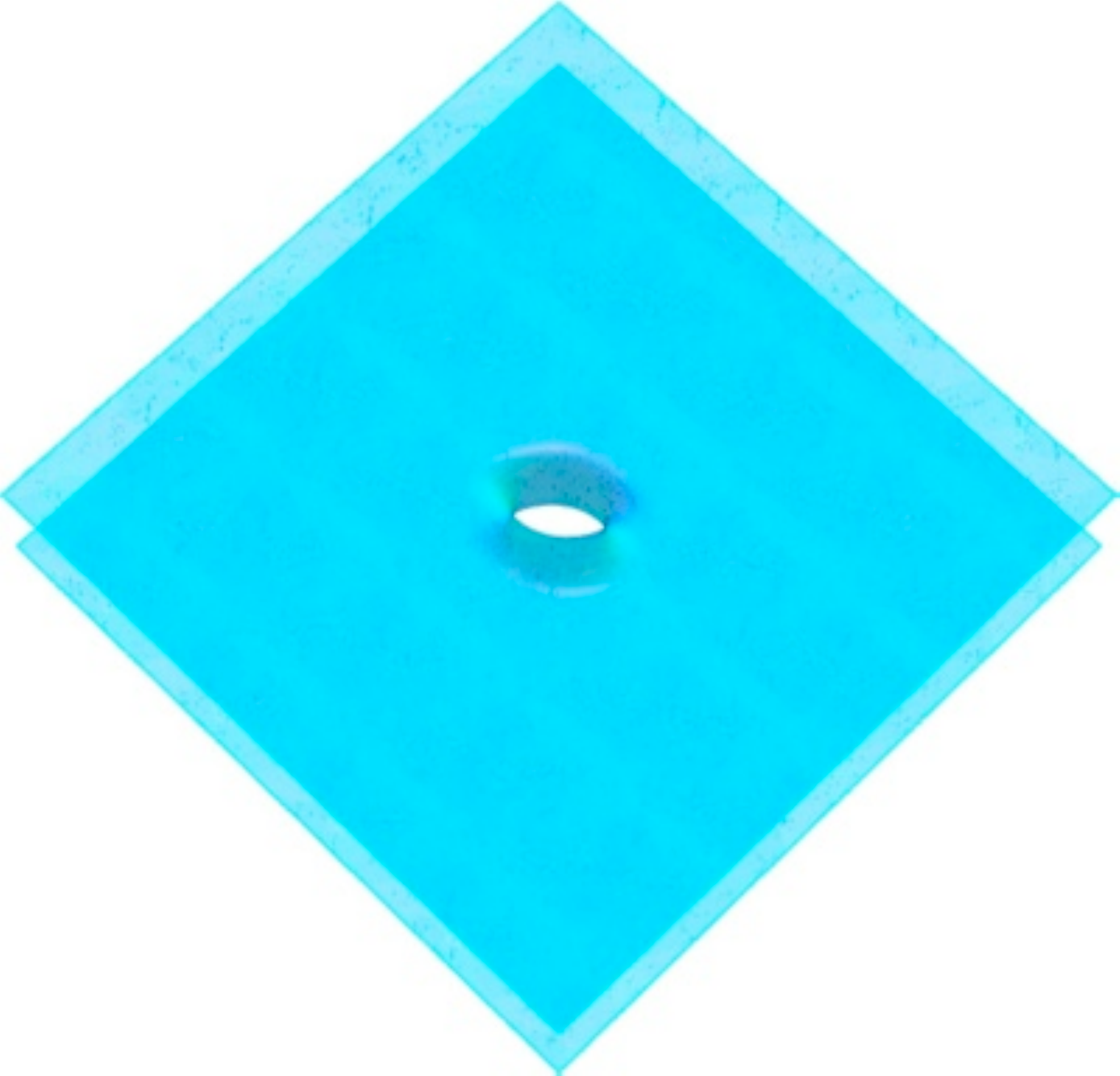}
\hspace{2cm}
\includegraphics[width=0.35\textwidth]{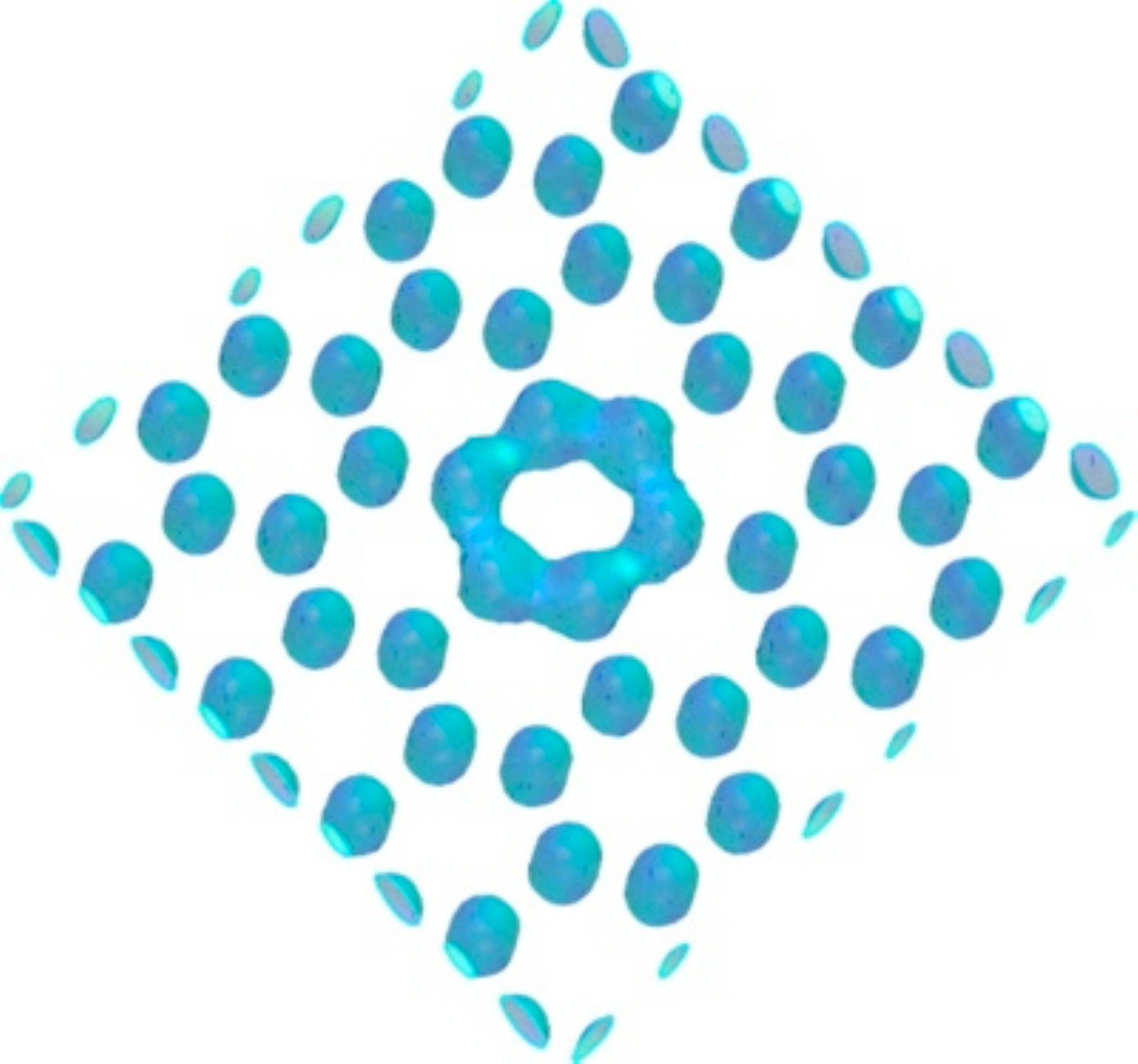}
\caption{The elementary  excitations above the extreme ground states illustrated in Fig.1 of the 
main part of the paper:
 a single plaquette flux in the `two-sheet' phase, and a single plaquette in the 'mulit-sphere' phase.}
\label{high_genus_exc}
\end{figure}

We briefly mention the elementary excitations of this model. 
In the `two-sheets' phase, 
which corresponds to couplings $\theta=0$ ($J_p=1$, $J_e=0$), 
the elementary excitation is a single plaquette with a $\tau$-flux giving rise to a
single `hole' as illustrated on the left in Fig.~\ref{high_genus_exc}.
These excitations are gapped with a gap size of $J_p$ and will delocalize
for small couplings  $J_e \neq 0$ forming quasiparticle bands. 
Similar to the ladder model the dispersion of this quasiparticle band can
be calculated perturbatively around the `two-sheets' limit.

In the opposite limit of `decoupled spheres',
which corresponds to couplings $\theta=\pi/2$ ($J_e=1$, $J_p=0$),
the elementary excitation is a `plaquette ring' where all edges around
a given plaquette have $\tau$-fluxes, as illustrated on the right in
Fig.~\ref{high_genus_exc}.
Again, a perturbative analysis allows to qualitatively describe the quasiparticle
band.

\begin{thebibliography}{10}
\bibitem{FootnoteModInv}
There exist in fact precisely two
CFTs of this central charge. Ours is the non-standard one whose spectrum
possesses certain two-fold degeneracies \cite{CappelliItzyksonZuber}.
\bibitem{RefFMatrix}
which is an extension (truncation) of the $6-j$- symbol of
$SU(2)$ to the quantum group $SU(2)_q$
for deformation parameter $q$ at root of unity \cite{QuantumSixJ-Reshetikhin}.
\bibitem{FootnotePlaquetteTerm}
The plaquette term is formulated in the same manner as the 
one in Eq.~\ref{plaq_term}.  The rung term is a projector of the fusion 
product of anyons $a_{i-1}$, $b_{i-1}$, as well as of anyons 
$a_{i+1}$ and $b_{i+1}$ onto the trivial particle.
\bibitem{FootnoteAdjacencyMatrix}
This is the matrix whose only non-vanishing  matrix elements  are 
$A_{\alpha,\alpha'}=1$ when $\alpha$ and $\alpha'$ are nearest neighbors
on the Dynkin diagram.
\end{thebibliography}
\end{document}